\documentclass[submission, Phys]{SciPost}
\hypersetup{
    colorlinks,
    linkcolor={red!50!black},
    citecolor={blue!50!black},
    urlcolor={blue!80!black}
}

\usepackage[bitstream-charter]{mathdesign}
\usepackage{physics}
\usepackage{xcolor}
\usepackage{subcaption}
\usepackage{chngcntr}

\DeclareSymbolFont{usualmathcal}{OMS}{cmsy}{m}{n}
\DeclareSymbolFontAlphabet{\mathcal}{usualmathcal}

\begin{document}

\begin{center}{\Large \textbf{\color{scipostdeepblue}{
Quantum complexity across thermal phase transition in the transverse field Ising chain with long-range couplings\\
}}}\end{center}

\begin{center}\textbf{
M. Adhikary\textsuperscript{1$\star$},
N. Ranabhat\textsuperscript{1,2} and
M. Collura\textsuperscript{1,3$\dagger$}
}\end{center}

\begin{center}
{\bf 1} International School for Advanced Studies (SISSA), via Bonomea 265, 34136 Trieste, Italy
\\
{\bf 2} University of Maryland, Baltimore County, Baltimore, Maryland, USA
\\
{\bf 3} INFN Sezione di Trieste, 34136 Trieste, Italy
\\[\baselineskip]
$\star$ \href{mailto:email1}{\small madhikar@sissa.it}\,,\quad
$\dagger$ \href{mailto:email2}{\small mcollura@sissa.it}
\end{center}

\section*{\color{scipostdeepblue}{Abstract}}
\textbf{\boldmath{%
We investigate the behavior of the Schmidt gap, the von Neumann entanglement entropy, and the non-stabiliserness in proximity to the classical phase transition of the one-dimensional long-range transverse-field Ising model (LRTFIM). Leveraging the time-dependent variational principle (TDVP) within a tensor-network formulation, we simulate thermal states through their purified tensor-network representations. Our results show that these observables—typically regarded as hallmarks of quantum criticality—exhibit pronounced and coherent signatures even at a classical thermal transition, highlighting the emergence of quantum complexity as the system nears thermal criticality.
}}

\vspace{\baselineskip}

\noindent\textcolor{white!90!black}{%
\fbox{\parbox{0.975\linewidth}{%
\textcolor{white!40!black}{\begin{tabular}{lr}%
  \begin{minipage}{0.6\textwidth}%
    {\small Copyright attribution to authors. \newline
    This work is a submission to SciPost Physics Lecture Notes. \newline
    License information to appear upon publication. \newline
    Publication information to appear upon publication.}
  \end{minipage} & \begin{minipage}{0.4\textwidth}
    {\small Received Date \newline Accepted Date \newline Published Date}%
  \end{minipage}
\end{tabular}}
}}
}



\vspace{10pt}
\noindent\rule{\textwidth}{1pt}
\tableofcontents
\noindent\rule{\textwidth}{1pt}
\vspace{10pt}

\section{Introduction}\label{sec:intro}

Understanding the interplay between classical statistical phenomena and quantum complexity is a fundamental challenge in many-body physics. In particular, classical thermal phase transitions, while traditionally described through macroscopic order parameters and equilibrium statistical mechanics \cite{landau1937theory,onuki2002phase}, can reveal subtle quantum signatures when examined through the lens of purified quantum states. This perspective is particularly relevant for systems where high control over individual constituents is possible, such as trapped-ion quantum simulators.\\

A paradigmatic platform for exploring such phenomena is the long-range one-dimensional transverse-field Ising model (LRTFIM) at finite temperature T \cite{Stinchcombe_1973, FisherPRL1972}. In this model, spins interact via a power-law exchange coupling that decays with distance as $d^{-\alpha}$, and are subject to a transverse magnetic field $h$. The LRTFIM exhibits a rich phase diagram in the temperature–field plane, with a quantum phase transition at $T = 0$ from a paramagnetic to a ferromagnetic phase \cite{Stinchcombe_1973}. The critical behavior of the model depends sensitively on the interaction exponent $\alpha$, which controls the decay of the exchange interactions, interpolating between a mean-field-like regime for small 
$\alpha$ and short-range interactions for large $\alpha$ \cite{FisherPRL1972,Gonzalez_Lazo_2021}. Beyond equilibrium, the model has attracted attention for its non-equilibrium dynamics, including interesting correlation propagation and entanglement growth~\cite{Hauke_2013, Schachenmayer_2013, Buyskikh_2016, Schneider_2021}, making it a versatile platform for quantum simulation, quantum information, and quantum computation studies~\cite{De_2025}.\\

To probe quantum aspects of thermal transitions in the LRTFIM, we employ tensor network methods, representing the thermal state in a purified form via the Matrix Product Density Operator (MPDO) ansatz \cite{PRLCirac2004, PRBsteven2005}. Purification allows any mixed state to be encoded as a pure state in an enlarged Hilbert space, introducing ancillary degrees of freedom that encode classical statistical mixing. In this framework, we can efficiently compute the entanglement of the purified state, the Schmidt gap of the entanglement spectrum, and measures of quantum complexity such as non-stabilizerness (``magic''). Importantly, these quantities capture genuine quantum correlations that are not apparent in the classical density matrix, highlighting how classical thermal states can encode nontrivial quantum features at criticality \cite{RMPvlatko2008, PRLHaldane2008, PRAmarioAlioscia2024}.\\

Non-stabilizerness, quantified via stabilizer R\'enyi entropies and related measures, provides a rigorous lens for characterizing the computational resources inherent in quantum states \cite{IOPVeitch_2014, AliosciaPRL2022}. While entanglement alone is insufficient to guarantee quantum computational advantage, stabilizer states being highly entangled yet classically simulable, non-stabilizerness captures the ``magic'' resources required for universal quantum computation \cite{IOPVeitch_2014, Pasquale_Calabrese_2004, PRDEsko2019, Annalen_der_Physik_Dalmonte_2022}. In our analysis, we evaluate these measures for finite-size systems and demonstrate that pronounced peaks in non-stabilizerness emerge at the thermal phase transition, signaling a sharp increase in quantum complexity precisely at criticality.\\

Methodologically, we obtain the purified states as Matrix Product States (MPS) using the Time-Dependent Variational Principle (TDVP) \cite{Ranabhat_2, lami2025beginnerslecturenotesquantum}. We focus on two representative values of the transverse field, $h = 0.0$ and $0.3$, and two interaction exponents, $\alpha = 0.8$ and $1.8$, while scanning the temperature to cross the paramagnetic-to-ferromagnetic transition. This procedure yields the MPS representation of the purified state at each temperature, enabling a direct evaluation of entanglement and magic.\\

In this work, we combine the LRTFIM at finite temperature, tensor network purification, and measures of entanglement and non-stabilizerness to provide a unified picture of how classical phase transitions can manifest quantum complexity. Our results demonstrate that even classically describable mixed states exhibit rich quantum structure in their purified representation, bridging classical statistical mechanics, quantum information theory, and experimental quantum simulation platforms. 

\section{Model}
\label{sec:intro}

The LRTFIM serves as a paradigmatic model for studying phase transitions and critical phenomena. It exhibits not only the ground state and finite-temperature phase transition but also the dynamical phase transition due to the emergent confinement phenomenon. The model Hamiltonian is given by,
\begin{align}
\begin{split}
    \hat{H} &= -\frac{1}{K_{\alpha}(N)}\sum_{i > j}\frac{|J|}{|i-j|^\alpha}s^z_{i}s^z_{j} - h\sum_i s^x_i
    \label{Eq:hamilTFIM}
\end{split}
\end{align}
Here, $s^{\nu}_i$ represents the spin one-half operator in $\nu = x,y,z$ direction at site $i$. The spins interact via a power-law potential that decays as the inverse of their separation distance raised to the power $\alpha$. The parameter interaction strength $|J|$ is set to unity and rescaled via the Kać normalization factor $K_{\alpha}(N)=1/(N-1)\sum_{i=1}^{N}(N-i)/i^\alpha$, to keep the energy per spin finite in thermodynamic limit for $\alpha<1$ scenario and just to put a energy rescaling for $\alpha>1$ case. The transverse magnetic field $h$ sets the strength of the off-diagonal term of the Hamiltonian responsible for the non-trivial dynamics. The parameter $\alpha$ tunes the range of spin-spin interaction in the model. At two extremes, the model becomes integrable and can be solved by exact analytical and numerical methods \cite{Defenu_2024, Ranabhat_1}. At $\alpha = \infty$, the model becomes the nearest-neighbor transverse field Ising model (TFIM), which can be effectively solved by mapping onto spinless fermions through the Jordan-Wigner transformation. This short-range universality class persists for $\alpha\geq3$. In the opposite limit,  $\alpha = 0$, the Hamiltonian can be written in terms of a collective sum of single operators, $S^{\nu} = \sum_i s^{\nu}$, and its thermodynamic limit properties are exactly captured by the mean field theory. More generally, the system belongs to the mean-field universality class for $\alpha\leq1.5$ ~\cite{FisherPRL1972}. \\

For strictly local interactions, one-dimensional models do not exhibit any finite-temperature phase transitions~\cite{Sachdev2011QPT}, as any ferromagnetic order is destroyed at \(T>0\). By contrast, long-range interactions increase the effective connectivity between spins, effectively raising their spatial dimension~\cite{eff_dim1,eff_dim2,eff_dim3,eff_dim4}. Quantitatively, a \(d\)-dimensional system with couplings decaying as \(J(r)\propto r^{-\alpha}\) displays universal behavior equivalent to a locally interacting system in an effective dimension  $d_{\text{eff}} = 2(d+z)/(\alpha-d)$, where \(z\) is the dynamical critical exponent~\cite{Defenu_2024, Dutta_2001, Monthus_2015}. Within the effective-dimension picture, lowering \(\alpha\) raises \(d_{\mathrm{eff}}\) across the thresholds separating short-range–like, nontrivial long-range, and mean-field regimes, providing a unified rationale for both the emergence of thermal order and the continuous evolution of critical exponents in the long-range transverse-field Ising model~\cite{Gonzalez_Lazo_2021}. The long-range power-law couplings render domain walls increasingly costly with their linear extent: the excess energy of a domain of length \(\ell\) scales as \(E_{\mathrm{DW}}(\ell)\sim \ell^{\,2-\alpha}\). For $1<\alpha<2$, this scaling overcomes the configurational entropy that drives kink proliferation in models with local interaction, thereby stabilizing long-range ferromagnetic order at finite temperature \cite{Ranabhat_1, Ranabhat_2, Ranabhat_3, Conf_1, Conf_2}. In one dimension, the mechanism yields a thermal transition for \(1<\alpha\le 2\) with a low temperature ferromagnetic phase transitioning to a high temperature paramagnetic phase, whereas for \(\alpha>2\) the model reverts to short-range behavior with no finite-\(T\) transition \cite{Knap_2013, Dutta_2001}. 

\section{Quantum-inspired probes for classical thermal transition}
\label{sec:3}
Traditionally, thermal phase transitions are identified by the expectation values of local order parameters and their susceptibilities, which change from nonzero to zero across a symmetry-breaking transition. For the LRTFIM, prior work has characterized the transition using such order-parameter diagnostics \cite{Gonzalez_Lazo_2021}. Here, we adopt a complementary, order-parameter-agnostic, quantum-information-inspired approach. Specifically, we construct the thermal Gibbs state in the locally purified tensor network representation as introduced by Werner et al. \cite{Werner_2016}, $\rho_{\beta} = X_{\beta}X_{\beta}^{\dagger}$. $X_{\beta}$ is called the purification operator and is represented as,

\begin{equation}
    [X_{\beta}]_{s_1,\ldots,s_i,\ldots,s_N}^{\kappa_1,\ldots,\kappa_i,\ldots,\kappa_N} = \sum_{\mu_1,\ldots,\mu_i,\ldots,\mu_{N-1}} \mathbb{A}_{\mu_1}^{s_1\kappa_1}\cdots\mathbb{A}_{\mu_{i-1},\mu_i}^{s_i\kappa_i}\cdots\mathbb{A}_{\mu_{N-1}}^{s_N\kappa_N}
    \label{Eq:Xbeta}
\end{equation}
where the physical indices $s_i$ and the Kraus indices $k_i$ lie in range $1\leq s_i \leq d$,  $1\leq \kappa_i \leq K$, respectively.  For simulating the thermal state, we keep the Kraus dimension equal to the physical dimension, $K = d=2$ (for spin-$\tfrac{1}{2}$). $\mu_i$ is the bond index and is in the range $1 \le \mu_i \le \mu_{\max}$ is the bond index with $\mu_{\max}$. $X_{\beta}$ is a matrix product state with each element being a rank-4 tensors $\mathbb{A}$. Equivalently, we can also represent $X_{\beta}$ in a vectorized form,

\begin{equation}
    |X_{\beta}\rangle = \sum_{s_1,\kappa_1,\ldots,s_i,\kappa_i,\ldots,s_N,\kappa_N} [X_{\beta}]_{s_1,\ldots,s_i,\ldots,s_N}^{\kappa_1,\ldots,\kappa_i,\ldots,\kappa_N} |s_1,\ldots,s_i,\ldots,s_N\rangle_{\text{phys}}\otimes|\kappa_1,\ldots,\kappa_i,\ldots,\kappa_N\rangle_{\text{aux}}.
\end{equation}

The thermal density matrix can then be constructed by a partial trace over the auxiliary space, $\rho_{\beta} = \text{Tr}_{\text{aux}}(|X_{\beta}\rangle\langle X_{\beta}|)$. This shows that $|X_{\beta}\rangle$ is a pure state defined in an enlarged Hilbert space of a physical system plus auxiliary (or Kraus) space, $\mathcal{H}_{\text{phys}}\otimes \mathcal{H}_{\text{aux}}$. For the details on the simulation of $|X_{\beta}\rangle$ by imaginary time TDVP, see Appendix \ref{App:0}. This representation guarantees the positivity of the density matrix by construction \cite{Werner_2016} and enables efficient evaluation of physical observables \cite{Ranabhat_2} like $\text{Tr}(\rho_{\beta}O) = \text{Tr}(X^{\dagger}_{\beta}OX_{\beta})$ as an expectation over pure state, $\text{Tr}(\rho_{\beta}O) = \langle X_{\beta}|O  \otimes \mathbb{1}_{\text{aux}}|X_{\beta}\rangle$. In this article, we will calculate two non-classical parameters directly over the purification $|X_{\beta}\rangle$ to study the thermal phase transition,

\begin{enumerate}
    \item the Schmidt gap of the bipartite entanglement spectrum, which captures changes in the structure of correlations across the transition
    \item the Stabilizer Rényi Entropy (SRE), which quantifies the deviation of the purification from stabilizer-like structure and thereby probes the complexity of the underlying mixed state.

\end{enumerate}

\subsection{Schmidt gap of entanglement spectrum}

 In the field of quantum many-body physics, Schmidt eigenvalues provide a crucial measure to characterize the entanglement spectrum. Given a pure state $\ket{X_\beta}$, these eigenvalues can be obtained for a given bipartition ( say $A$ and $B$ ) of the system through Schmidt decomposition, $\ket{X_\beta}=\sum_i\lambda_i\ket{i_A}\otimes\ket{i_B}$, where $\{\lambda_i\}$'s are the Schmidt coefficients and the $\sum_i$ runs over the nonzero terms. The gap between the first two largest coefficients, known as the  Schmidt gap (SG), is considered an important order parameter to identify quantum phase transition \cite{PRLSanpera, PRBIlluminati2013, PRBSenpera2013, PRBGabriele, PRBPollman, PRBAbolfazl, NcomStasinska2014}.\\
 
 The ground state study of TFIM reveals a transition of SG from $0$ to $1$ as $h$ changes from $0$ to a higher value \cite{PRBGabriele}. SG taking the value $1$ is equivalent to one of the Schmidt vectors being dominant, and for TFIM, this implies that the system belongs to a product state. Whereas a $0$ value of $SG$ for a lower value of $h$ implies the presence of the GHZ-like state ( Greenberger-Horne-Zeilinger ), i.e., a mixture of $|11\cdots1> $ and $|00\cdots0>$. Such interesting features of SG motivate us to investigate its behavior near the classical phase transition of LRTFIM.\\
 
 In the present work, we consider the central partition of a pure state MPS $\ket{X_\beta}$ for the open spin chain of length $N$. The central partition is a reliable measure of the maximum entanglement attained by the system.  Thereafter, we calculate $\Delta$, defined as the difference between the squares of the two largest Schmidt eigenvalues of the entanglement spectrum, together with its susceptibility as a function of $\beta$,
    
 \begin{align}
     \begin{split}
       \Delta&=\lambda_1^2-\lambda_2^2  \\ 
       \chi&=\partial\Delta/\partial \beta
     \end{split}
     \label{Eq:SG and chai}
 \end{align}

Another interesting quantity related to the Schmidt coefficients is the Von Neumann entanglement entropy (S) of the complete system, defined as,
\begin{equation}
     S=-\sum_i \lambda_i^2 \log_2\lambda_i^2
    \label{Eq:entropy}
\end{equation}

This is a well-studied measure of quantum correlation for a pure state. At zero temperature, as $h$ goes from $0$ to $1$, $S/N$ goes from $1/N$ to $0$ showing a singularity at the criticality, \cite{PRAjulien2005,PRpappalardi2024}. The nature of this divergence depends upon $\alpha$. To analyze the quantum complexity of the model near the classical critical point, we calculate $\Delta,~ \chi,~S$ for system sizes $N=16,32,64,100,250,300,350,400$. We discuss the results in Sec .~\ref {sec:4}.

\subsection{Stabilizer Rényi Entropy (SRE)}
\label{sec:3.2}
Nonstabilizerness or quantum magic, $M$ is considered to be a good measure of constructional complexity of a quantum state from a quantum technological perspective \cite{AliosciaPRL2022,Leone_2024}. It gives us a deep insight into how hard it is to prepare the corresponding state in a classical computer. For a stabilizer state, $M$ is always zero, thus $M$ behaves in a different way than the standard entanglement. A non-zero $M$ mirrors the quantum complexity in the state. Motivated by such an interesting character of $M$, in the second part of our analysis, we focus on estimating the quantum magic $M$ by means of calculating the n-SRE of the pure state MPS found as a result of TDVP. We follow the perfect sampling of the Pauli string framework to calculate $M_n$, as prescribed in ~\cite{LamiMario2023,Haug2023stabilizerentropies}. For a $N-$qubit system in a pure state $\rho=\ket{\psi}\bra{\psi}$, the $n-$ SRE is defined as following~\cite{AliosciaPRL2022},
\begin{equation}
    M_n(\rho)=\frac{1}{1-n}\log \sum_{\sigma \in \mathcal{P}_N}\frac{1}{2^N}Tr[\rho\boldsymbol{\sigma}]^{2n}
    \label{Eq:n-SRE}
\end{equation}

Here, $\boldsymbol{\sigma}$  denotes a string of $N$ Pauli operators, $\boldsymbol{\sigma}=\prod_{i=1}^N\sigma_i$. Each $\sigma_i$ can take four values $\{\sigma^0,\sigma^1,\sigma^2,\sigma^3\}$. The set $\mathcal{P}_N$contains all possible such strings, and therefore has the size $4^N$. For a certain pure state $\ket{\psi}$, we construct the corresponding pure state density matrix $\rho$. If we regard the positive real valued function, $\boldsymbol{\prod}_\rho(\boldsymbol{\sigma})=(1/2^N)Tr[\rho\boldsymbol{\sigma}]^{2}$, as a probability distribution over $\boldsymbol{\sigma}$,  Eq.~\ref{Eq:n-SRE} takes the form of a standard Rényi entropy . This interpretation is valid since $\sum_{\boldsymbol{\sigma\in\mathcal{P}_N}}\boldsymbol{\prod}_\rho(\boldsymbol{\sigma})=1$ \cite{AliosciaPRL2022}. With this consideration, n-SRE calculation turns out to be a computation of the average of $(n-1)^{\text{th}}$ power of the distribution function itself. With this simplification, Eq.~\ref{Eq:n-SRE} can be written as $M_n(\rho)=1/(1-n)\log \overline{\boldsymbol{\prod}_\rho(\boldsymbol{\sigma})^{n-1}}-N\log2$, for $n>1$. For $n=1$ the expression reduces to Shannon entropy, $M_1(\rho)=-\overline{\log \boldsymbol{\prod}_\rho(\boldsymbol{\sigma})}-N\log2$. The key point to calculate the average quantities involved in the expressions of $M_n$ is to perform a perfect sampling of the Pauli string $\boldsymbol{\sigma}$ with probability $\boldsymbol{\Pi}_{\rho}(\boldsymbol{\sigma})$ following the method mentioned in ~\cite{LamiMario2023}. \\

In our present case, instead of a ground state MPS, we deal with $\ket{X_\beta}$ of a doubled Hilbert space, as discussed before. To mirror the effect of the enlarged Hilbert space, we double the Pauli string by considering $\boldsymbol{\sigma\tilde{\sigma}}$, which is an operator sequence of length $2N$. In Appendix~\ref{App:C} we discuss the challenges in using Eq.~\ref{Eq:n-SRE} without any Pauli string enlargement. In this version, the enlarged operator string is defined as, $\boldsymbol{\sigma}\boldsymbol{\tilde{\sigma}}=\prod_{i=1}^N\sigma_i\tilde{\sigma}_i$. At each site $i$, there are two Pauli operators $\sigma_i$ and $\tilde{\sigma}_i$, where the latter acts on the auxiliary leg. Thus here, Eq.~\ref{Eq:n-SRE} takes the following form,

\begin{equation}
    M_n(\rho)=\frac{1}{1-n}\log \sum_{\boldsymbol{\sigma\tilde{\sigma}} \in \tilde{\mathcal{P}}_N}\frac{1}{4^N}Tr[\rho\boldsymbol{\sigma\tilde{\sigma}}]^{2n}
    \label{Eq:n-SRE_MPDO}
\end{equation}

 The string operator,  $\boldsymbol{\sigma\tilde{\sigma}} \in \tilde{\mathcal{P}}_N$ contains now $4^{2N}=16^N$ possibilities. The important point is that even in this case the sampling prescription is valid as $\boldsymbol{\prod}_\rho(\boldsymbol{\sigma} \boldsymbol{\tilde{\sigma}})=(1/4^N)Tr[\rho\boldsymbol{\sigma} \boldsymbol{\tilde{\sigma}}]^{2}$ remains a positive real valued function and $\sum_{\boldsymbol{\sigma\tilde{\sigma}} \in \tilde{\mathcal{P}}_N}\boldsymbol{\prod}_\rho(\boldsymbol{\sigma} \boldsymbol{\tilde{\sigma}})=1$  and therefore can be treated as a distribution function of $\boldsymbol{\sigma} \boldsymbol{\tilde{\sigma}}$. See Appendix~\ref{App: A} for technical details on performing the conditional sampling for the extended Pauli string operator. The expressions for SRE becomes $M_1(\rho)=-q_1-N\log4$ and $M_n=\frac{1}{(1-n)}\log q_n-N\log4$ for $n>1$, 
 where we introduce $q_1=\overline{\log\boldsymbol{\Pi}_{\rho}}
 =1/\mathcal{N}\sum_{i=1}^{\mathcal{N}}\log\boldsymbol{\Pi}_{\rho}({\boldsymbol{\sigma}_i\tilde{\boldsymbol{\sigma}}}_i)$ and  $q_n=\overline{\boldsymbol{\Pi}_{\rho}^{(n-1)}}=1/\mathcal{N} \sum_{i=1}^{\mathcal{N}}\boldsymbol{\Pi}_{\rho}({\boldsymbol{\sigma}_i\tilde{\boldsymbol{\sigma}}}_i)^{(n-1)}$,
 and where in the sums the strings $\boldsymbol{\sigma_i\tilde{\sigma_i}}$ have been extracted according to the probability $\boldsymbol{\Pi}_{\rho}({\boldsymbol{\sigma}_i\tilde{\boldsymbol{\sigma}}}_i)$.
 We thus compute the magic density $ m_n = M_n/N$, and evaluate its statistical uncertainty using the expression below (which captures that bias introduced by the logarithm for $n\neq 1$), 
 \begin{align}
\begin{split}
    \delta m_1&= \sqrt{\text{var}[\log\boldsymbol{\Pi}_\rho]/\mathcal{N}}\\
    \delta m_2&=\sqrt{(1/\mathcal{N}) \text{var}[\boldsymbol{\Pi}_\rho]/q_2^2}
\end{split}
\label{EQ:err in m}
\end{align}
Here, `var' stands for the variance. It is important to notice that the expression for  $\delta m_2$ contains a ratio of two fluctuating variables $\text{var}[\boldsymbol{\Pi}_\rho]/q_2^2$, unlike $\delta m_1$. Thus, for the same sample size $\mathcal{N}$, $\delta m_2$ appears larger than $\delta m_1$. In this part of our study we focus on the 3 lowest system sizes, which are $N=16,~32,~64$. In section ~\ref{sec:4.2} we show how  $m_n$ changes with inverse temperature $\beta$ for a fixed $\alpha,~h$ and $n=1,2$.  
\begin{figure*}[t!]
    \centering
    \begin{subfigure}[b]{\textwidth}
        \includegraphics[width=1.\textwidth]{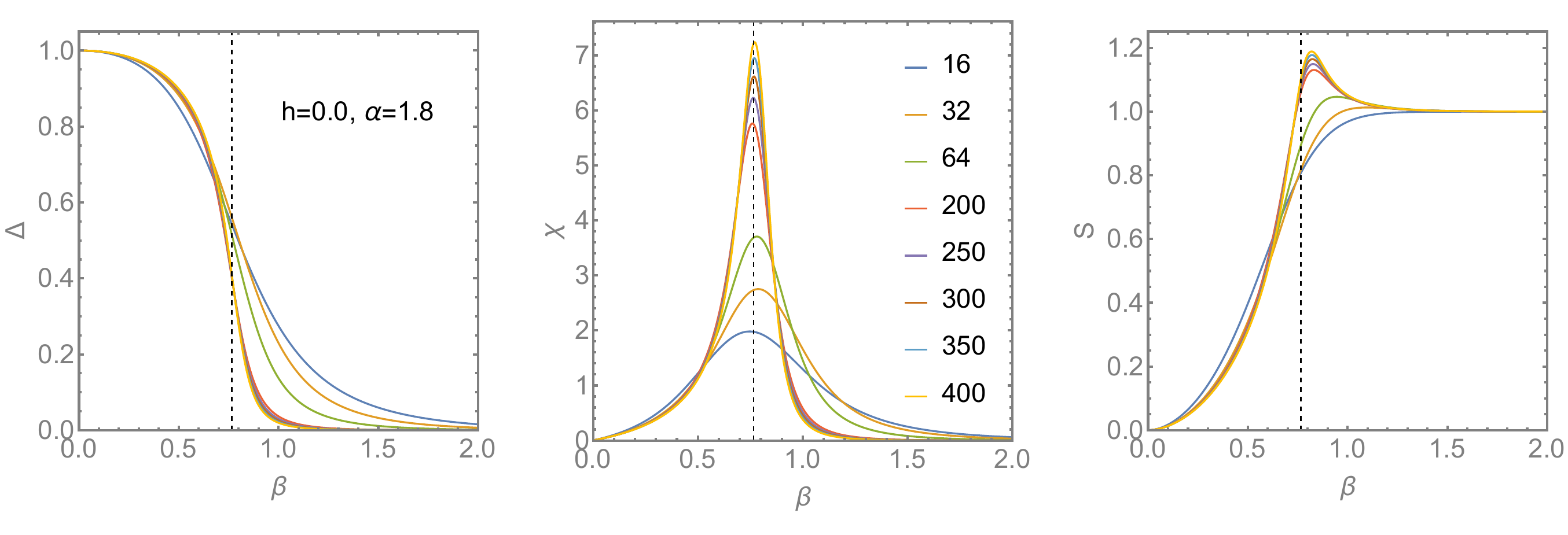}
        \caption{}
        \label{fig:gap_vs_beta_diffN_a}
    \end{subfigure}
    \hfill
    \begin{subfigure}[b]{\textwidth}
        \includegraphics[width=\textwidth]{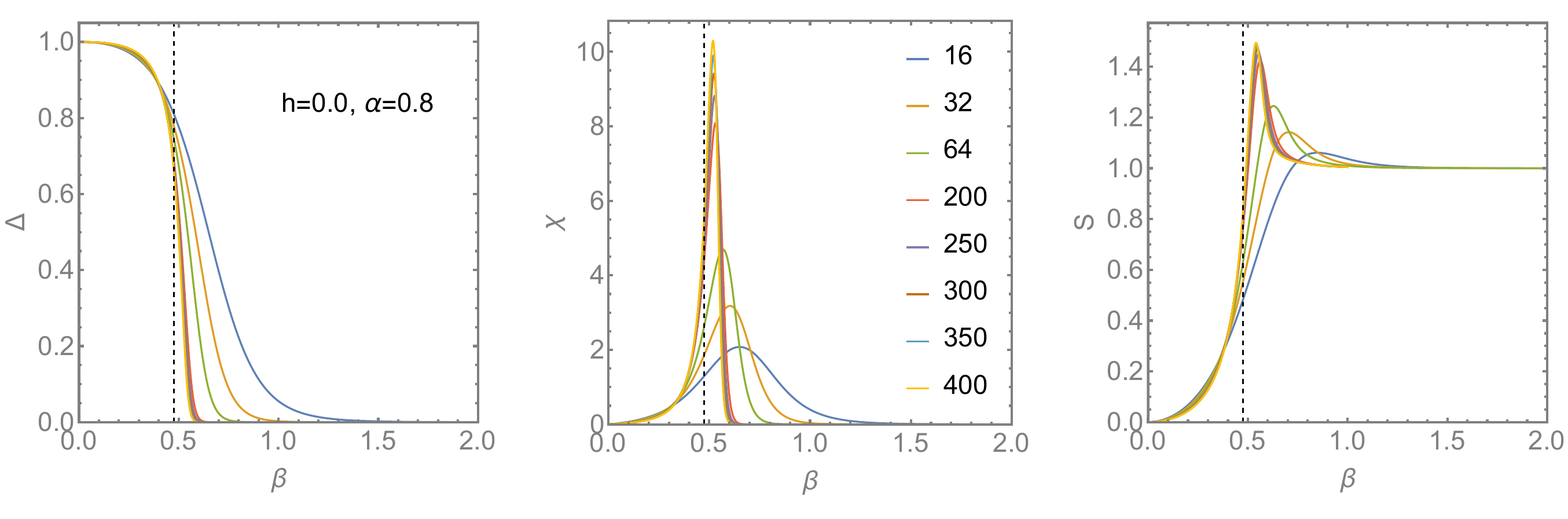}
        \caption{}
        \label{fig:gap_vs_beta_diffN_b}
    \end{subfigure}    
    \caption{Evolution of the eigen value gap $\Delta$, the corresponding susceptibility $\chi$, and the von Neumann entanglement entropy $S$ of central bipartition, as functions of $\beta$ for  $h = 0.0$, system sizes $N=16,32,64,200,250,300,350,400$ and (a) $\alpha = 1.8$; (b)$\alpha = 0.8$. The vertical dashed lines in all the plots are our estimate of the corresponding $\beta_c$'s from FSS analysis.}
    \label{fig:gap_vs_beta_diffN}
\end{figure*}

\begin{figure*}[]
    \centering
    \begin{subfigure}[b]{\textwidth}
        \includegraphics[width=1.\textwidth]{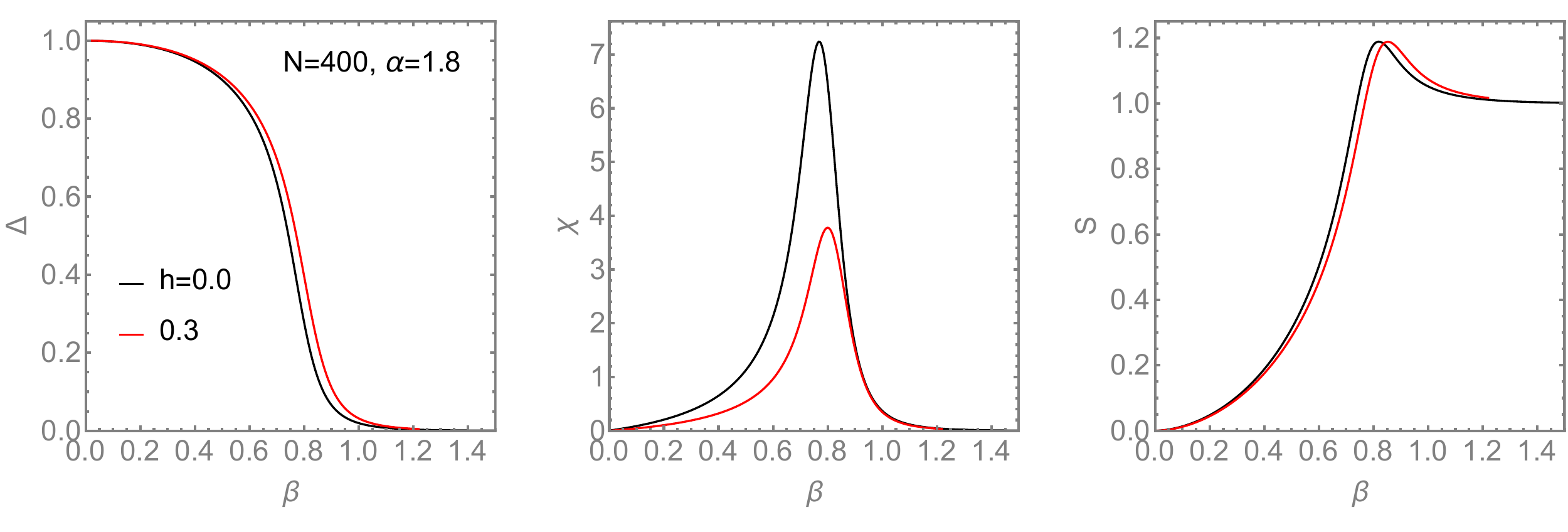}
        \caption{}
        \label{fig:gap_vs_beta_1_a}
    \end{subfigure}
    \hfill
    \begin{subfigure}[b]{\textwidth}
        \includegraphics[width=\textwidth]{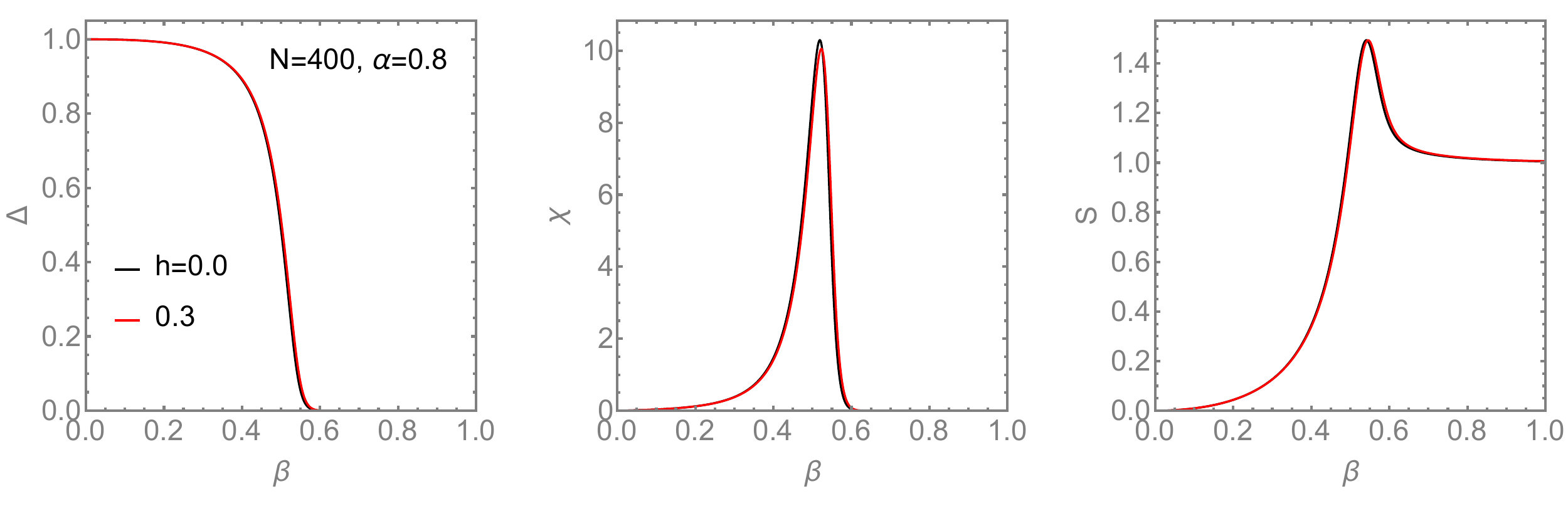}
        \caption{}
        \label{fig:gap_vs_beta_1_b}
    \end{subfigure}
    \hfill
    \begin{subfigure}[b]{\textwidth}
        \includegraphics[width=\textwidth]{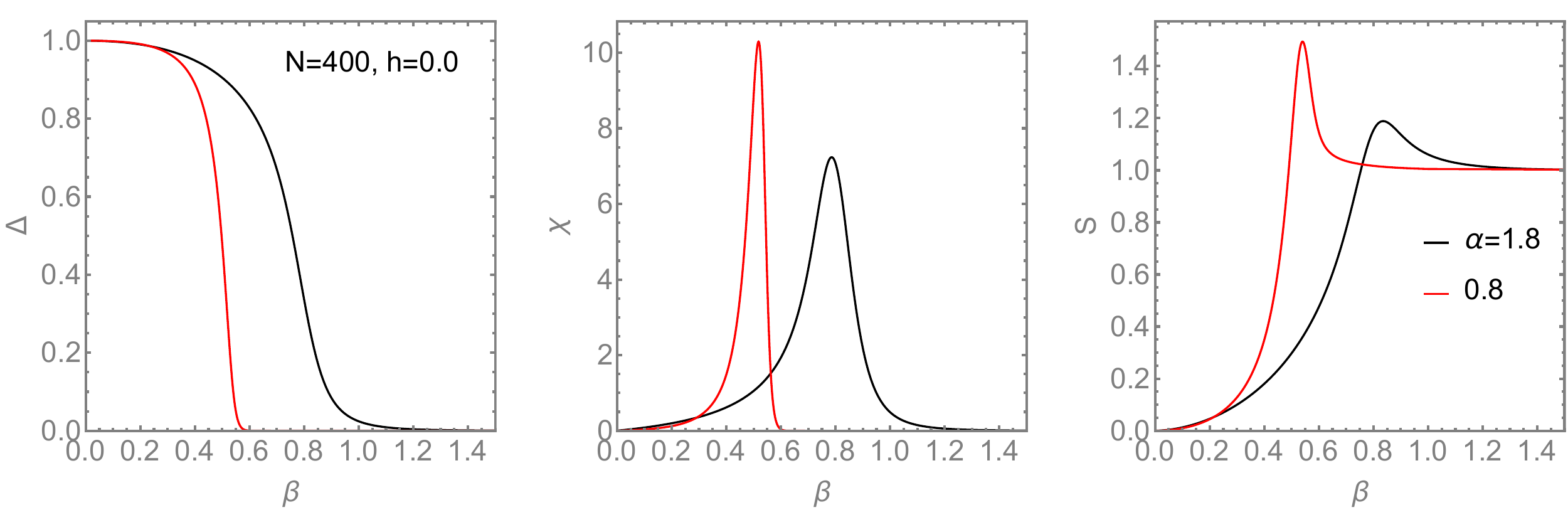}
        \caption{}
        \label{fig:gap_vs_beta_1_c}
    \end{subfigure}
    \caption{Comparison of the $\beta$-dependence of $\Delta$,  $\chi$ and  $S$, for fixed $N=400$: (a) $h=0.0,~0.3$ for $\alpha=1.8$; (b) $h=0.0,~0.3$ for $\alpha=0.8$; (c) $\alpha=1.8,~0.8$ for $h=0.0$.} 
    \label{fig:gap_vs_beta_1}
\end{figure*}

\section{Results and Discussion}
\label{sec:4}

In Fig.~\ref{fig:gap_vs_beta_diffN}, we show the evolution of $\Delta$, $S$, and $\chi$ as functions of the inverse temperature $\beta$ for eight different system sizes $N$, while keeping $\alpha = 1.8$ in Fig.~\ref{fig:gap_vs_beta_diffN_a} and $\alpha = 1.8$ in Fig.~\ref{fig:gap_vs_beta_diffN_b}, with $h = 0.0$. Focusing on the leftmost panel, which displays $\Delta$ versus $\beta$, we observe that at high temperatures (low $\beta$), $\Delta \approx 1$. This corresponds to $\lambda_1 \approx 1$, indicating the absence of quantum entanglement and the classical nature of the thermal state. In this regime, thermal fluctuations dominate and effectively suppress quantum correlations. As $\beta$ increases, $\Delta$ decreases smoothly from $1$ to $0$, with a more rapid decay near the critical point, as evidenced by the $\chi = \partial \Delta / \partial \beta$ plot. At large $\beta$, $\Delta \approx 0$ signifies the entangled ground state given by an equal superposition of the all-up and all-down spin configurations of the LRTFIM at $h=0$.
The two rightmost panels of Fig.~\ref{fig:gap_vs_beta_diffN} show the entanglement entropy $S$ as a function of $\beta$. Starting from $ S\approx 0$ at high temperature, $S$ increases with decreasing temperature, exhibiting a peak before saturating to $1$ (note that $S$ is scaled by $\log 2$). This behavior reflects the transition from a maximally mixed, entanglement-less thermal state at high $T$ to a GHZ-like entangled state at low $T$. The collective temperature dependence of $\Delta$, $\chi$, and $S$ thus provides clear signatures of a finite-temperature transition. Fig~\ref{fig:gap_vs_beta_diffN} further illustrates the finite-size effects in these quantities near the transition. We find that the critical inverse temperature $\beta_c$ is more strongly affected by finite-size effects for $\alpha = 0.8$. In this case, the pick in $\chi$ or $S$ gradually shifts towards lower $\beta$ as $N$ increases. On the other hand, for $\alpha=1.8$, there is no noticeable shift in these picks with increasing $N$. The result consistently captures the fact that the finite-size effect is more pronounced in the long-range limit than in the short-range regime.  In Sec .~\ref {Sec:4.1} we analyze the $\beta_c$'s through a finite-size scaling of the order parameter $\Delta$. Our estimate of $\beta_c$'s for both the $\alpha$'s are marked through the vertical dashed lines.\\

In Fig.~\ref{fig:gap_vs_beta_1_a} and \ref{fig:gap_vs_beta_1_b}, we show the variation of the three quantities $\Delta$, $S$, and $\chi$ for two different values of the transverse field $h=0.0,0.3$, with $N=400$ and $\alpha=1.8$. As $h$ increases from $0.0$ to $0.3$, the critical features in all three observables shift to larger values of $\beta$. Fig.~\ref{fig:gap_vs_beta_1_a} shows that the shift is prominent for $\alpha=1.8$. In contrast, for $\alpha=0.8$ we find the shift is less noticeable. Quantitatively, the displacement is approximately $\Delta \beta \sim 0.06$ and $\sim 0.003$, indicating that the critical temperature decreases more for the short range regime  when a nonzero field is applied. See Table~\ref{tab:crititable}.
This behavior is consistent with the physical expectation that, in the presence of a transverse field, a smaller amount of thermal fluctuation is sufficient to drive the system into the paramagnetic phase~\cite{Dutta_2001, Gonzalez_Lazo_2021}. Consequently, the phase transition occurs at a lower temperature for larger magnetic fields, eventually terminating at a quantum critical point located at $h = h_c$ at zero temperature. A quick DMRG check yields $h_c \approx 1.55(2)$ and $\approx 1.85(2)$ for $\alpha=1.8$ and $0.8$ respectively. This explains why the observed shift in the critical region between $h=0.0$ and $h=0.3$ remains smaller for the mean-field limit than for the shorter-range limit. Since the numerical cost of the full tensor-network simulations increases substantially with larger $h$, we restrict our analysis to these two field values. In Appendix~\ref{App:D}, we provide the corresponding DMRG results for both values of $\alpha$. Similarly, in Fig.~\ref{fig:gap_vs_beta_1_c} we compare the evolution of $\Delta$, $\chi$, and $S$ as functions of $\beta$ for $\alpha = 1.8$ and $\alpha = 0.8$, keeping $N = 400$ and $h = 0.0$ fixed. In this case, we observe a noticeable shift of the critical point toward lower values of $\beta$ as $\alpha$ decreases. This substantial displacement when moving from $\alpha = 1.8$, which lies in the intermediate regime between short- and long-range interactions, to $\alpha = 0.8$, which is deep in the mean-field-like regime, is consistent with earlier findings~\cite{Gonzalez_Lazo_2021}.
Next, we perform a finite-size scaling analysis of these three order parameters to estimate the critical inverse temperature $\beta_c$ for $\alpha = 0.8, 1.8$ and $h = 0.0, 0.3$.

\begin{figure*}[t!]
    \centering
    \begin{subfigure}[t]{0.5\textwidth}
        \centering
        \includegraphics[width=0.9 \textwidth]{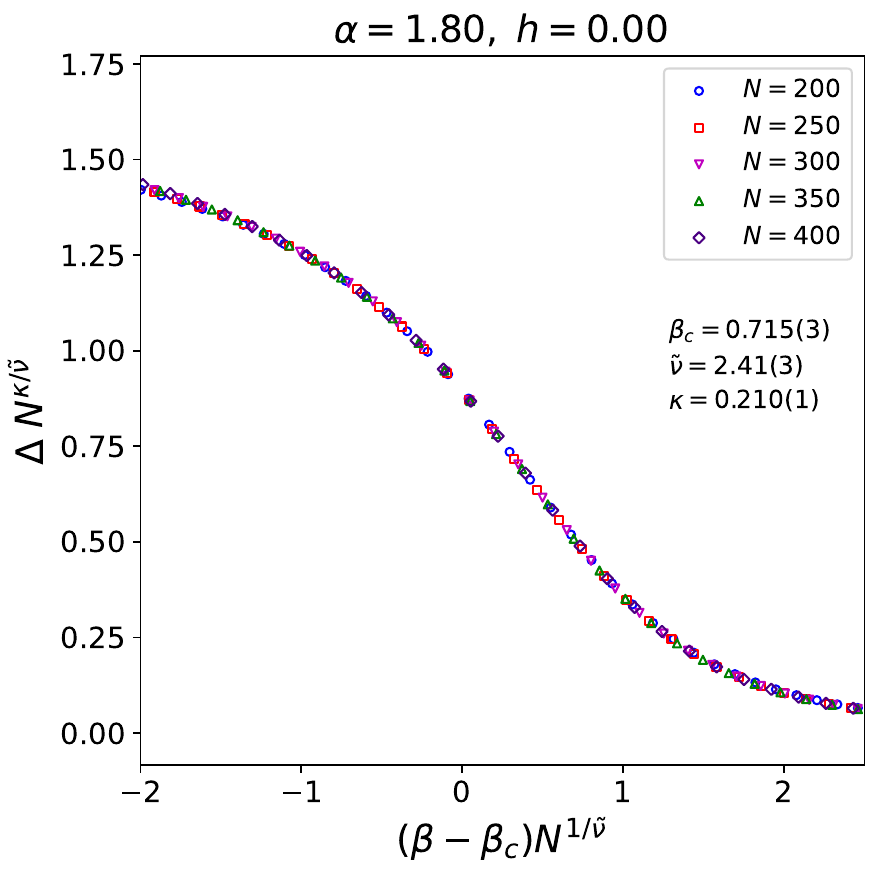}
        \caption{}
        \label{fig:scaling1}
    \end{subfigure}%
    ~ 
    \begin{subfigure}[t]{0.5\textwidth}
        \centering
        \includegraphics[width=0.9\textwidth]{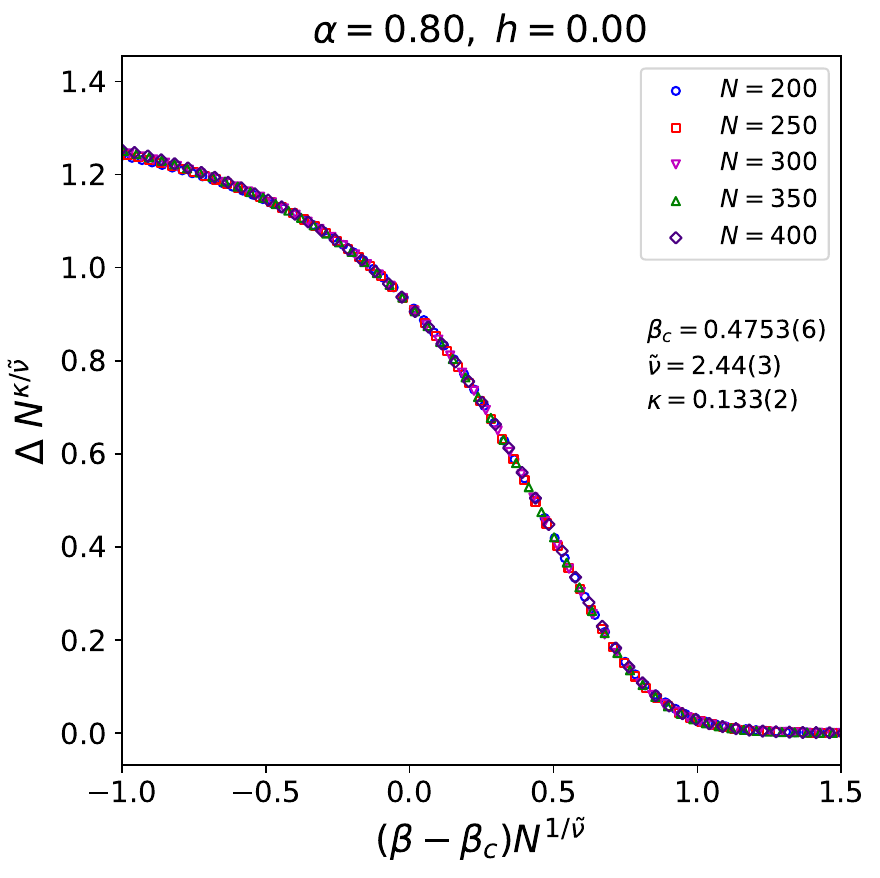}
        \caption{}
        \label{fig:scaling}
    \end{subfigure}
    \caption{Data collapse of $\Delta$ at $h = 0.0$ for (a) $\alpha = 1.8$ and (b) $\alpha = 0.8$, shown for five different system sizes. }
    \label{fig:scaling}
\end{figure*}

\subsection{Finite size scaling of Schmidt gap}
\label{Sec:4.1}

To analyze the finite temperature critical behavior, we perform a finite-size scaling (FSS) analysis following the standard prescription for classical order parameters ~\cite{Binder1987, Kadanoff_1966, FSSsandvik1997}. The Schmidt gap $\Delta$ is a good observable in the study of quantum criticality and is known to exhibit critical finite-size scaling~\cite{PRLSanpera, Torlai2018}. If it is also a good indicator of thermal phase transition, then near the critical point $\beta_c$, $\Delta$ is expected to obey the universal scaling. According to the recent approach for thermal phase transition, known as Q finite-size scaling (Q-FSS) ~\cite{Weigel2015,Kenna_2014,BERCHE2012115}, this universal scaling can be written as, 
\begin{equation}
\Delta(N,\beta) = N^{-\kappa/\tilde{\nu}}\,\mathcal{F}\!\left[(\beta-\beta_c)\,N^{1/\tilde{\nu}}\right],
\end{equation}
 Here $\mathcal{F}$ is the universal critical scaling function. $\tilde{\nu}=\text{max}(1, d/d_{uc})\nu$ and $\nu$ is the correlation length critical exponent.  $\kappa$ is a scaling exponent associated with the Schmidt gap. The quantity $d$ is the physical dimension of the system, which is $1$ in our case, and $d_{uc}$ is the upper critical dimension of the system and is $d_{uc}=2(\alpha-1)$ for $\alpha\in[1,3]$. As proposed in the framework of Q-FSS,
\begin{align}
    \begin{split}
        \tilde{\nu} &= \nu ~~ \text{for} ~~ d\leq d_{uc} ~~(\alpha \geq1.5 )\\
        &=\frac{d}{d_{uc}}\nu ~~\text{for}~~ d> d_{uc} ~~ (\alpha<1.5)
    \end{split}
    \label{Eq:rescalesnu}
\end{align}
For $\alpha\leq1.5$, the model belongs to the mean field critical behavior, and here the renormalization group studies have shown $\nu=1/(\alpha-1)$ ~\cite{FisherPRL1972}. For the intermediate $1.5<\alpha\leq 2.0$, $\nu$ continuously varies with $\alpha$ in a nontrivial way~ \cite{Weigel2015,Parisi2014}, as quantum fluctuation plays an important role here. \\

The rescaled Schmidt gap $\Delta N^{\kappa/\tilde{\nu}}$, when plotted as a function of $(\beta-\beta_c)N^{1/\tilde{\nu}}$, should collapse onto a single universal curve for different system sizes $N$. In practice, we treat the critical exponents $\kappa$ and $\tilde{\nu}$, together with the critical inverse temperature $\beta_c$, as free fitting parameters to obtain the best data collapse. These exponents provide a quantitative characterization of the critical behavior. We restrict the FSS analysis to system sizes $N=200,250,300,350,400$ and estimate $\{\beta_c, \kappa, \tilde{\nu}\}$ for the four combination of $\{\alpha,h\}$ as summarized in the table below

\begin{table}[h]
    \centering
    \begin{tabular}{||c|c|c|c|c||}
\hline
      $\alpha$ & h & $\beta_c$ & $\kappa$ & $\tilde{\nu}$\\
      \hline
     1.8 & 0.0 & 0.715(3) &0.210(1)&2.41(3)\\
     \hline
     1.8 & 0.3 & 0.8020(6)&0.270(3)&2.34(3)\\
     \hline       
     0.8 & 0.0 & 0.4753(6) &0.133(2)&2.44(3)\\ 
     \hline
     0.8 & 0.3 & 0.4779(4)&0.133(2)&2.47(4)\\
      \hline
\end{tabular}
    \caption{The list of critical exponents and inverse critical temperatures for four different combinations of $\{\alpha,h\}$.}
    \label{tab:crititable}
\end{table}

In Fig.~\ref{fig:scaling}(a) and (b), we present the resulting data collapse for 
$\alpha = 1.8$ and $\alpha = 0.8$, respectively, at $h=0.0$. 
We observe that 
for a fixed field $h$,  $\beta_c$ is systematically smaller for 
$\alpha = 0.8$ than for $\alpha = 1.8$. Conversely, for fixed $\alpha$, increasing 
the field from $h = 0.0$ to $h = 0.3$ shifts $\beta_c$ to larger values. It is also relevant to mention that the same field effects $\beta_c$ more in the short-range limit than in the long range limit.
For zero field $\alpha=1.8$ the system is in the intermediate regime and our data collapse result estimates $\tilde{\nu}=\nu=2.41(3)$. This estimate agrees with the finding of Ricci-Tersenghi \emph{et al.} \cite{Parisi2014} where they calculate critical exponents for thermal phase transition for $\alpha=1.875,1.654$ and find $\nu\approx2.425$ and $\approx1.976$, respectively by means of Monte Carlo simulation. Our result $2.41(3)$ for $\alpha=1.8$ falls between the two $\nu$'s that they estimate and lies close to their result for $\alpha=1.875$. We also compare our result with Troung \emph{et al.} \cite{Troung1999} and again find good agreement. Therefore, we can conclude that the Schmidt gap is good enough to capture the thermal critical behavior and is capable of indicating that the system belongs to the Ising universality class. 
For $\alpha = 0.8$, the system lies deep inside the long range regime, and as mentioned before, for this the system should fall under the mean-field universality class. Here, the extraction of $\nu$ from $\tilde{\nu}$ needs a non-trivial rescaling, and it is less well characterized in the literature for $\alpha < 1$ ~\cite{Gonzalez_Lazo_2021}. Our scaling analysis yields  $\tilde{\nu}=2.44(3)$; however, the limited theoretical understanding in this regime restricts a direct quantitative comparison with established results.

\subsection{Behavior of the upper bound of magic near criticality}
\label{sec:4.2}

In this part, we analyze the thermal phase transition through the behavior of $m_n$ as a function of $\beta$. As discussed in Sec .~\ref{sec:3.2}, the fact that each tensor in the MPS representation is a rank-4 effectively doubles the system size for the purposes of computing the $n$-SRE, making the calculation equivalent to treating a system of size $2N$. This imposes practical limitations on the choice of $N$, since larger physical system sizes require substantially more samples $\mathcal{N}$ to achieve reliable statistical accuracy, as indicated by Eq.~\ref{EQ:err in m}.
Consequently, in this part of our analysis, we restrict ourselves to comparatively smaller system sizes for evaluating the non-stabilizerness densities $m_n$. First we discuss the key features for $N=32$. In the subsequent subsection we compare the data for $N = 16, 32, 64$.  For all results presented below, we fix the number of samples to $\mathcal{N} = 10^4$.\\

\begin{figure*}[t!]
    \centering
    \includegraphics[width=1.\textwidth]{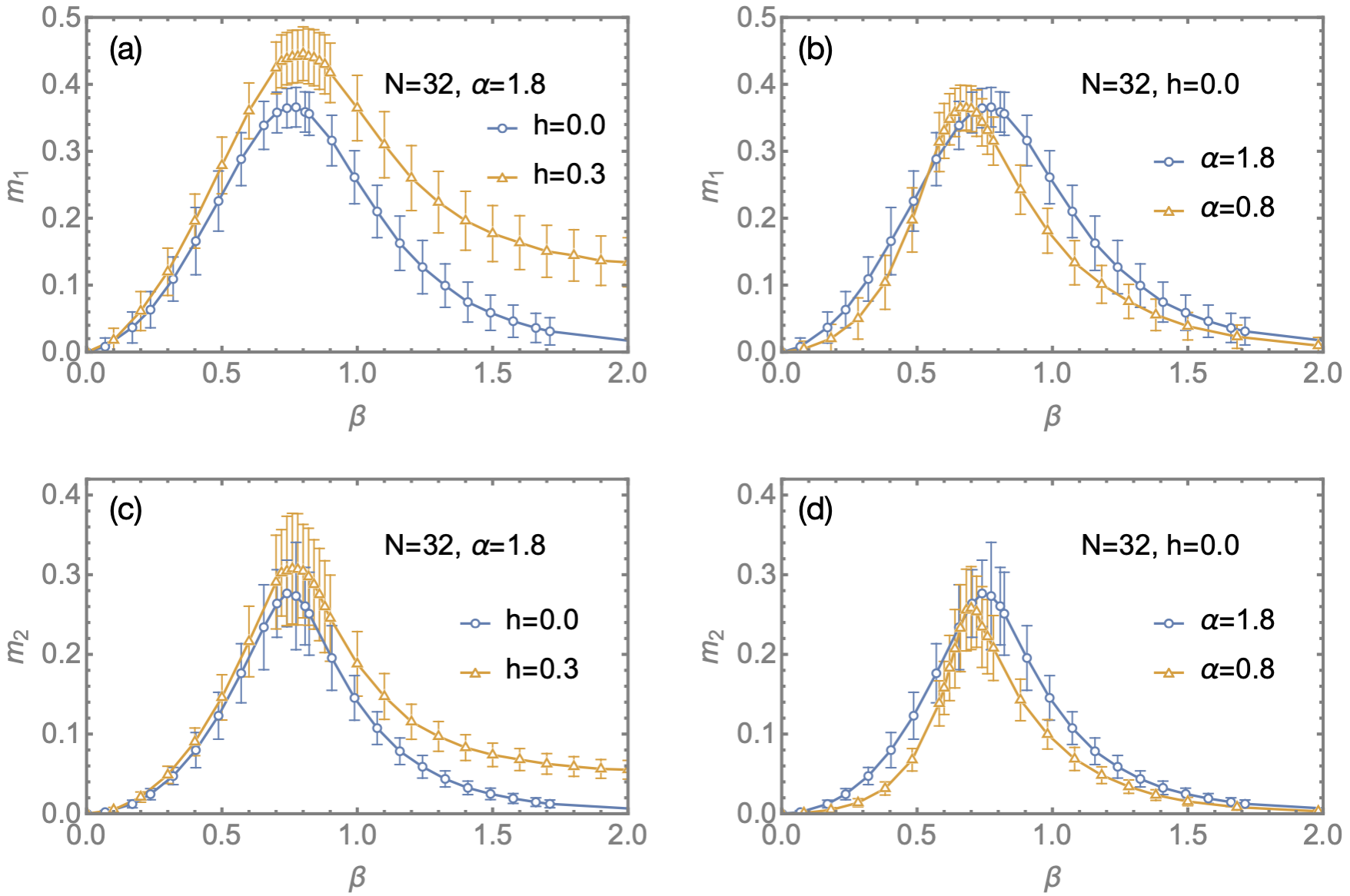}
    \caption{Evolution of the nonstabilizerness density, $m_n$ as a function of $\beta$ . Panel (a) and (b) shows $m_1$ versus $\beta$: - In (a) $\alpha=1.8$; the blue and yellow lines correspond to $h=0.0$ and $h=0.3$  respectively. - in (b) $h=0.0$: the blue and yellow lines correspond to $\alpha=1.8,~0.8$ respectively. Panels (c) and (d) are the same as (a) and (b) but show the variation of $m_2$ instead of $m_1$. All plots are for $N=32$ .}
    \label{fig:m_vs_beta}
\end{figure*}

Fig.~\ref{fig:m_vs_beta}(a) shows the evolution of $m_1$ for $h = 0.0$ and $h = 0.3$ (blue and orange curves, respectively) at fixed $\alpha = 1.8$ and $N = 32$. Similarly, Fig.~\ref{fig:m_vs_beta}(b) displays the corresponding behavior for $\alpha = 0.8$ and $\alpha = 1.8$ (blue and orange curves, respectively), with $h = 0.0$ and $N = 32$. Figures~\ref{fig:m_vs_beta}(c) and (d) present the same information for $m_2$. At fixed $\alpha$, the peak position of $m_{1,2}$ exhibits only a weak dependence on $h$ in Fig.~\ref{fig:m_vs_beta}(a) and (c), whereas the magnitude of $m_{1,2}$ increases clearly with $h$. In contrast, Fig.~\ref{fig:m_vs_beta}(b) and (d) show a more pronounced shift of peaks toward lower $\beta$ with nearly identical amplitudes,  as $\alpha$ is reduced from $1.8$ to $0.8$, reflecting the distinct universality classes associated with these interaction exponents. These trends are consistent with the behavior of $\Delta$, $\chi$, and $S$ discussed earlier.
In the zero-field case, for both choices of $\alpha$, $m_{1,2}$ vanishes at small $\beta$, then increases as $\beta$ grows, reaches a maximum, and eventually returns to zero as $\beta \rightarrow \infty$. This vanishing of $m_{1,2}$ in both the $\beta \to 0$ and $\beta \to \infty$ limits is expected for the LRTFIM at zero field: the system is maximally mixed at $\beta = 0$ and becomes ferromagnetically ordered at very large $\beta$, and both limits correspond to stabilizer states with zero magic. For $h = 0.3$, however, $m_1$ ($m_2$) saturates to $0.1341 \pm 0.04$ ($0.055 \pm 0.012$) at large $\beta$, as any nonzero field introduces non-stabilizerness into the system.
The qualitative features of $m_1$ and $m_2$ as functions of $\beta$ are similar; however, the peak of $m_2$ is lower and narrower. The statistical error in $m_{1,2}$ is proportional to $m_{1,2}$ itself and therefore vanishes when the magic is zero. Near the peak, $\delta m_1 \sim 8\%$ for $h=0.0$, increasing to $\sim 9\%$ for $h=0.3$. For $m_2$, $\delta m_2$ reaches $\sim 14\%$ at zero field and increases to $\sim 22\%$ at $h = 0.3$ for $\mathcal{N}=10^4$ samples. As discussed in Sec .~\ref{sec:3.2}, the maximum $\delta m_2$ is significantly larger than the maximum $\delta m_1$.\\

As discussed in Sec.~\ref{sec:3.2}, the quantity 
$m_n$ that we compute in this work is not the absolute or “true” magic of the thermal state at inverse temperature $\beta$. Rather, it corresponds to a particular choice of purification or embedding of the mixed state into a larger pure-state system. In principle, a more rigorous definition of magic for a mixed state would require a more involved minimization procedure (for example, across different purification schemes)~\cite{Leone_2024, Warmuz_2025}. Because we are not performing this full optimization, our estimate of magic is an upper bound — the actual magic could be strictly lower. Nonetheless, even within our framework, the signal we obtain appears strong enough to reliably detect the thermal phase transition.

\begin{figure*}[t]
    \centering
    \includegraphics[width=1.\textwidth]{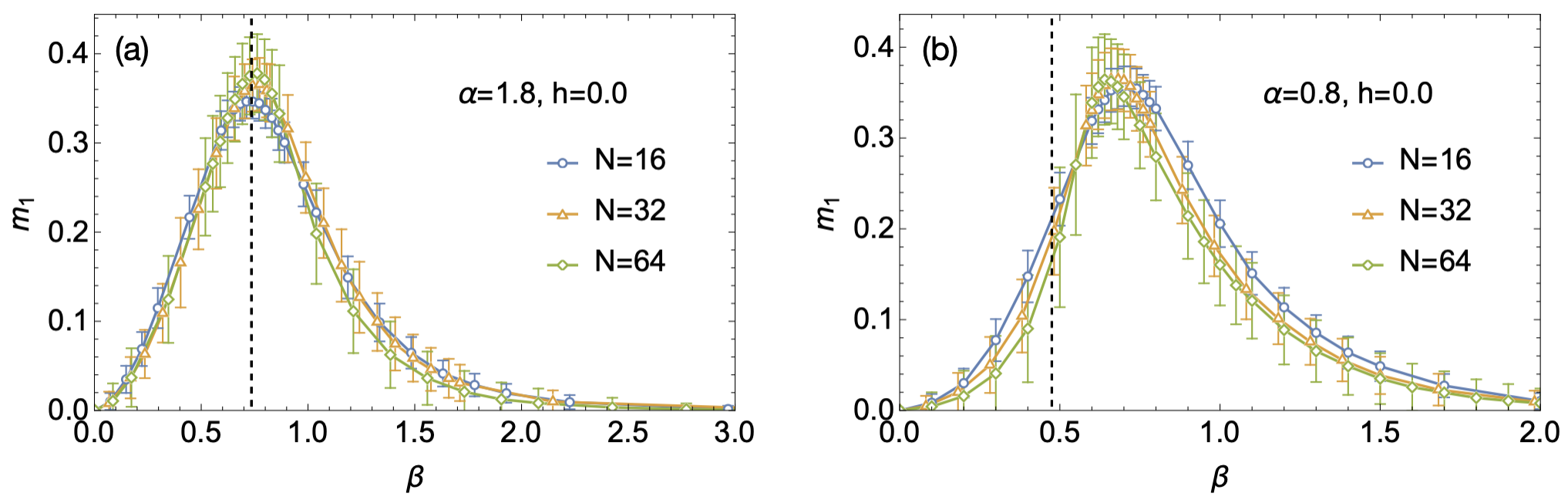}
    \caption{Evolution of the nonstabilizerness density, $m_n$ as a function of $\beta$ for $N=16$ (Blue lines) and $N=32$ (Orange lines), and $N=64$ (Green line), in the absence of an external field. (a) and (b) shows $m_1$ versus $\beta$ for $\alpha=1.8,0.8$ respectively. The black dashed line in both plots indicates $\beta_c$ that we estimate from FSS .}
    \label{fig:m_vs_beta_diff_N}
\end{figure*}

\subsection{Finite size effect on magic}
In  Fig.\ref{fig:m_vs_beta_diff_N} we compare the $m_1$ results for three different system sizes $N=16,32,64$ for both $\alpha=1.8,0.8$. The vertical dashed lines here refer to the $\beta_c$ that we extract from FSS of the Schmidt gap data for larger $N$'s in Sec.\ref{Sec:4.1}. Interestingly, for $1.8$, the system-size effect is not very prominent. Thus, the signal from the $m_1$ even for a lower $N$ is capable of giving a good indication of the critical point. Whereas for $0.8$ the estimate of $\beta_c$ from FSS is significantly away from the peaks of the three $m_1$ curves. Noticeably, the peak shifts towards the dashed line as $N$ increases from $16$ to $64$. Therefore, the system size effect is larger in $\alpha=0.8$ than in $\alpha=1.8$. This is again consistent with LRTFIM being in the long-range regime for the first value of $\alpha$, whereas being in the short range  regime for the other. Here we do not show the system size effect for $m_2$, as for all sampling purposes we kept $\mathcal{N}=10^4,$ which is not good enough for $N=64$ in order to get a lower variance.

\section{Conclusion and outlook}

In this work, we investigated the classical thermal phase transition of the one-dimensional spin-$1/2$ long-range transverse-field Ising model (LRTFIM), examining two representative regimes: the long-range interacting case with $\alpha = 0.8$ and the more short-range-like case with $\alpha = 1.8$. Using the TDVP algorithm within the tensor-network framework, 
we approximated thermal states through their purified MPS representations and analyzed both the Schmidt gap, capturing quantum entanglement, and the stabilizer Rényi entropies (SREs), probing quantum magic and numerical complexity.\\

For the entanglement study, we considered eight different system sizes  and extracted critical temperatures and critical exponents via a finite-size scaling analysis of the Schmidt gap for the five largest systems ($N = 200, 250, 300, 350, 400$).  Our estimate of $\nu$ seems quite consistent with previous work, at least in the short range limit. For the magic analysis, we restricted ourselves to the three smaller system sizes ($N = 16, 32, 64$) due to the computational cost of evaluating SREs. Remarkably, both indicators independently signal the thermal transition: the Schmidt gap 
collapses from $1$ to $0$, while the magic displays a pronounced peak as a function of $\beta$. For $\alpha = 1.8$, the transition points inferred from these two diagnostics agree well; for $\alpha = 0.8$, quantitative differences appear, reflecting stronger finite-size effects in the long-range regime. This distinction between mild and strong size dependence provides a consistent picture of the crossover between short-range and long-range behavior.\\

It is important to emphasize that, for the specific class of purification we employ, our analysis does not yield the \emph{true} magic or entanglement of the thermal mixed state: both quantities are affected by the MPS purification ansatz and by the finite bond dimension. Nevertheless, the approximation is sufficiently robust for the purpose of locating the thermal transition, which is the primary objective of our study.\\

Several natural directions emerge from our work. First, one may explore alternative tensor-network-based purification, such as non-locally purified tensor networks, isometric MPO encoding, or hybrid Clifford-assisted constructions. Different purification schemes may reduce the bias associated with the chosen ansatz, potentially yielding more accurate estimates of magic for mixed states and improving numerical stability near criticality.\\

Second, it would be interesting to investigate whether SREs for mixed states can be computed directly within an MPDO representation, without relying on a fixed purification. Although it remains unclear whether SREs for mixed states constitute a fully faithful measure of magic, they still represent a meaningful indicator, as they quantify the entropy of the state’s probability distribution in the Pauli basis. Developing efficient contraction or sampling strategies for MPDO-based SREs (as preliminarily depicted in the Appendix~\ref{App:C}) could significantly extend the applicability of magic diagnostics to larger system sizes or to real-time dynamics.\\

Finally, our results highlight the potential of complexity-based markers, such as magic and features of the entanglement spectrum, as complementary probes of phase transitions in long-range quantum systems. Extending this analysis to other models, interaction exponents, or dynamical settings (e.g., quenches, Floquet protocols, or monitored dynamics) may yield 
new insights into the interplay between long-range interactions, classical criticality, and quantum complexity.
 
\section*{Acknowledgments}
The authors acknowledge valuable discussions with Alessio Lerose, Luca Tagliacozzo, Emanuele Tirrito, Nicol\`o Defenu and Andrea Trombettoni.
M. A. and M. C. acknowledge support from the PNRR MUR project PE0000023-NQSTI. The numerical simulations for this project were performed on the Ulysses v2 cluster at SISSA and on the Leonardo cluster at CINECA, through the SISSA–CINECA agreement for access to high-performance computing resources.

\appendix
\numberwithin{equation}{section}

\section{Finite temperature density operator}
\label{App:0}
The finite-temperature states can be simulated by casting the density operator as locally
purified tensors \cite{PRLCirac2004,Werner_2016}. The thermal density operator is defined by the Gibbs distribution
$\hat{\rho}_\beta = \frac{e^{-\beta \hat{H}}}{\mathrm{Tr}[e^{-\beta \hat{H}}]}$. At $\beta = 0$ (infinite temperature),
The state is maximally mixed and can be written as a tensor product of local identities,
$\hat{\rho}_0 = \bigotimes_{i=1}^N \mathbb{1}_{s_i}$
where each $\mathbb{1}_{s_i}$ is a unit matrix of size $(d, d)$, i.e.
$\mathbb{1}_{s_i} = [\delta_{s_i,s_i'}]_{d\times d}$,
and $d$ is the dimension of the local Hilbert space. The density operator for any finite temperature (non-zero $\beta$) can be expressed as
\begin{subequations}
\begin{align}
\hat{\rho}_\beta &\propto e^{-\beta \hat{H}} \\
                 &= e^{-\frac{\beta}{2}\hat{H}}\, \hat{\rho}_0\, e^{-\frac{\beta}{2}\hat{H}}.
\end{align}
\end{subequations}

We keep the density operator in a locally purified form $\hat{\rho} = X X^\dagger$ at each stage, where $X$ is represented as a tensor as defined in Eq.~\ref{Eq:Xbeta}.

The density operator initialized at infinite temperature can be purified to a finite temperature
in trotterized steps:
\begin{subequations}
\begin{align}
\hat{\rho}_{\beta + d\beta}
  &= e^{-\frac{d\beta}{2}\hat{H}} \hat{\rho}_\beta e^{-\frac{d\beta}{2}\hat{H}} \\
  &= e^{-\frac{d\beta}{2}\hat{H}} X X^\dagger e^{-\frac{d\beta}{2}\hat{H}} \\
  &= \left( e^{-\frac{d\beta}{2}\hat{H}} X \right)
     \left( e^{-\frac{d\beta}{2}\hat{H}} X \right)^\dagger.
\end{align}
\label{Eq:rho}
\end{subequations}

Eq.~\ref{Eq:rho} can be simulated using imaginary-time TDVP
($-i\,dt \rightarrow -d\beta$), evolving only one half of the locally purified operator $X$
and never contracting $X$ with $X^\dagger$ during the evolution, thereby strictly preserving
the locally purified structure.
\begin{figure}[h]
    \centering
    \includegraphics[width=0.5\textwidth]{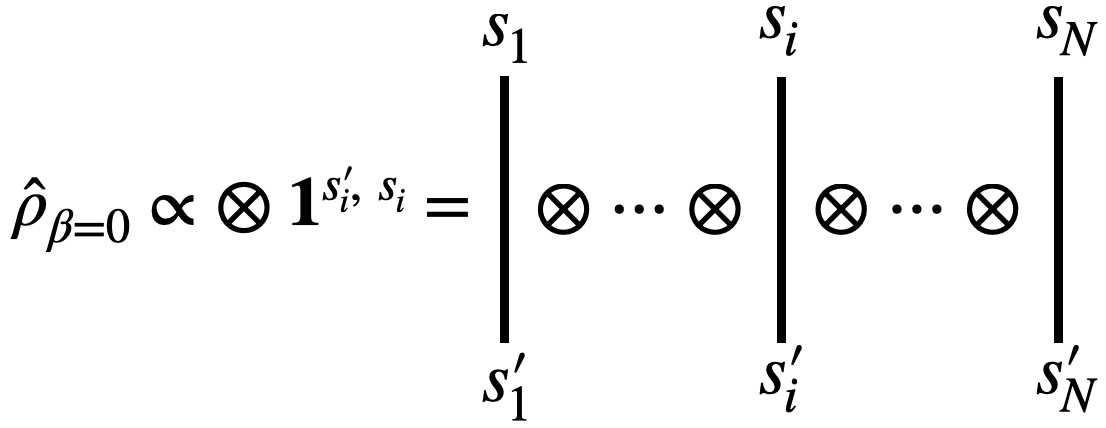}
    \caption{Maximally mixed density operator at $\beta=0$ as the tensor product of the local identities.}
    \label{fig:diag1}
\end{figure}

\begin{figure}[h]
    \centering
    \includegraphics[width=0.8\textwidth]{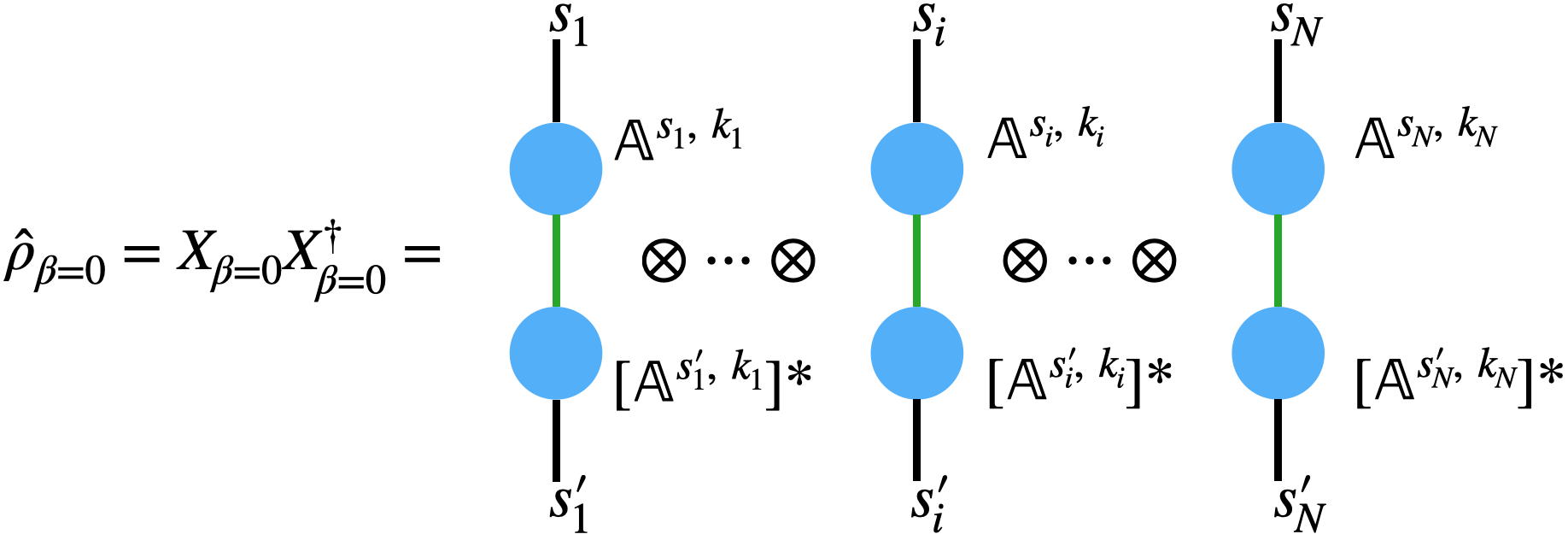}
    \caption{ Representing $\hat{\rho}_{\beta=0}$ in the locally purified form.}
    \label{fig:diag2}
\end{figure}

Fig.~\ref{fig:diag1} shows the tensor notation of the infinite temperature density operator $\hat{\rho}_{\beta=0}$, which is a tensor product of identity matrices of size $(d, d)$. Rather than working with the density operator as an MPO, we represent the density operator in the
locally purified form \cite{Werner_2016, IOPJaschke_2019}, which is positive semi-definite by construction, and keep it in locally purified form at every stage of the thermal purification process. In Fig.~\ref{fig:diag2} we represent $\hat{\rho}_{\beta=0}$ in the locally purified form $X_{\beta=0}X^{\dagger}_{\beta=0}$, where the index in green is an auxiliary index called the Kraus index.

\begin{figure}[h]
    \centering
    \includegraphics[width=.6\textwidth]{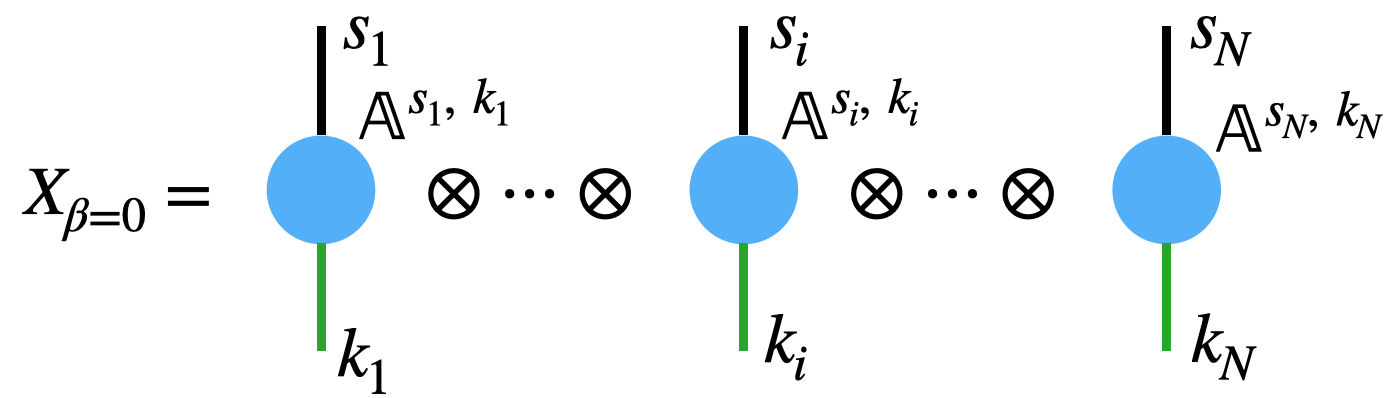}
    \caption{  One half of the $\hat{\rho}_{\beta=0}$ in the locally purified form.}
    \label{fig:diag3}
\end{figure}

We can now evolve one of the halves ( $X$or $X^\dagger$) as shown in Eq.~\ref{Eq:rho}, and the evolution on the other half is its trivial conjugate. This approach is computationally efficient as we can work with cheaper MPS instead of more expensive MPDO. In Fig.~\ref{fig:diag3}, one half of the $\hat{\rho}_{\beta=0}$ in locally purified form is shown; from here on, we will only work with this half.\\

\begin{figure}[h]
    \centering
    \includegraphics[width=.34\textwidth]{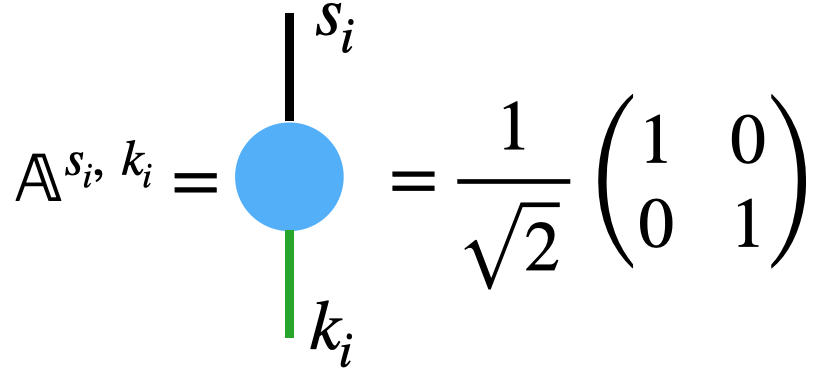}
    \caption{Choice of $\mathbb{A}^{s_i,~k_i}$ to preserve the trace of $\hat{\rho}$.}
    \label{fig:diag4}
\end{figure}

Algebraically, $X_{\beta=0}$ can be written as,
\begin{equation}
X_{s_1,k_1,\ldots,s_N,k_N}
= X^{s_1,k_1} \cdots X^{s_N,k_N}.
\end{equation}
For a system of spin-$\tfrac{1}{2}$ particles, we choose
\begin{equation}
X^{s_i,k_i} = \frac{1}{\sqrt{2}}
\begin{pmatrix}
1 & 0 \\
0 & 1
\end{pmatrix}, \quad i \in \{1,2,\ldots,N\},
\end{equation}
as shown in Fig.~\ref{fig:diag4}. This particular choice is taken to preserve the trace of the density operator,

\begin{equation}
\sum_k \mathbb{A}^{s,~k} (\mathbb{A}^{s',~k})^\dagger
= \frac{1}{2}
\begin{pmatrix}
1 & 0 \\
0 & 1
\end{pmatrix}.
\end{equation}

\begin{figure}[h]
    \centering
    \includegraphics[width=.6\textwidth]{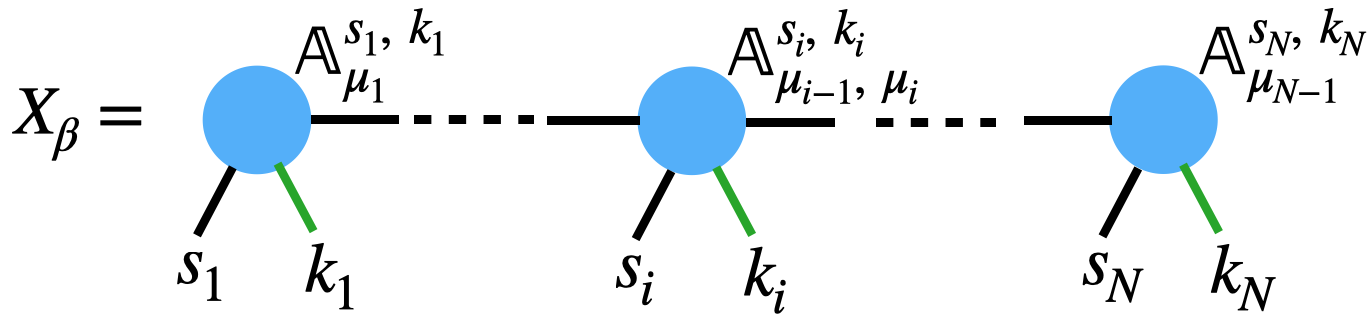}
    \caption{ $X_{\beta}$ in MPS form.}
    \label{fig:diag5}
\end{figure}

Finally, we reshape $X_{\beta=0}$ from a sequence of $2\times2$ matrices to
a sequence of four-legged tensors of shape $(1,2,2,1)$ as shown in Fig.~\ref{fig:diag5}, forming an MPS of bond dimension~1. The finite-temperature density operator is then obtained by iteratively applying the imaginary-time TDVP evolution in Eq.~\ref{Eq:rho}.
This procedure allows the simulation of thermal states efficiently using matrix-product representations while maintaining positivity by construction.\\

\begin{figure}[h]
    \centering
    \includegraphics[width=.7\textwidth]{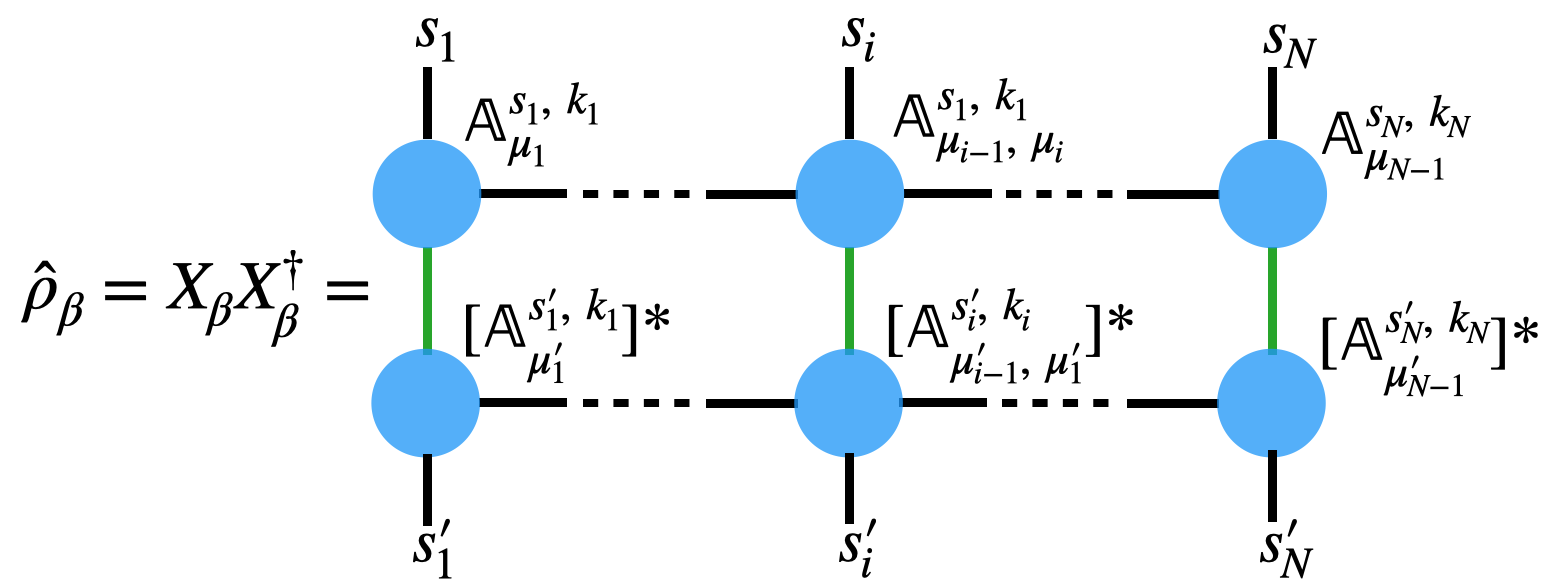}
    \caption{$\hat{\rho}_{\beta}$ in terms of $X_{\beta}$.}
    \label{fig:diag6}
\end{figure}

Now that we have our initial state as an MPS, we can simulate a finite-temperature density operator $\hat{\rho}_{\beta}$ by solving Eq.~\ref{Eq:rho}.
We employ two-site imaginary time TDVP with discrete Trotter time steps $dt = 0.001$ to simulate the thermal states. Unlike real-time evolution, the bond dimension does not grow exponentially and is maximum around criticality but remains well below the maximum bond dimension $\mu_{\text{max}} = 256$. At every bond, we truncate all the singular values less than $10^{-8}$.

\section{Conditional sampling for the enlarged Hilbert space}

\label{App: A}
 The complete probability to construct a certain $\boldsymbol{\sigma} \boldsymbol{\tilde{\sigma}}\in\tilde{\mathcal{P}}_N$, can be written in terms of $N$ conditional probabilities in the following way,

\begin{equation}
    \boldsymbol{\prod}_\rho(\boldsymbol{\sigma} \boldsymbol{\tilde{\sigma}})= \pi_{\rho}(\sigma_1\tilde{\sigma}_1)\pi_{\rho}(\sigma_2\tilde{\sigma}_2|\sigma_1\tilde{\sigma}_1)\cdots\pi_{\rho} (\sigma_i\tilde{\sigma}_i|\sigma_1\tilde{\sigma}_1 \cdots \sigma_{i-1}\tilde{\sigma}_{i-1})\cdots
    \label{Eq:condi_prob}
\end{equation}
Here,
\begin{align}
    \begin{split}
        & \pi_{\rho}(\sigma_i\tilde{\sigma}_i|\sigma_1\tilde{\sigma}_1 \cdots \sigma_{i-1}\tilde{\sigma}_{i-1}) =\pi_{\rho}(\sigma_i\tilde{\sigma}_i)/\pi_{\rho}(\sigma_1\tilde{\sigma}_1 \cdots \sigma_{i-1}\tilde{\sigma}_{i-1})\\
        & \pi_{\rho}(\sigma_1\tilde{\sigma}_1 \cdots \sigma_{i}\tilde{\sigma}_{i}) = \frac{1}{4^N}\sum_{\boldsymbol{\sigma\tilde{\sigma}} \in \tilde{\mathcal{P}}_{N-i}} \Tr(\rho \sigma_1\tilde{\sigma}_1 \cdots \sigma_{i}\tilde{\sigma}_{i}\boldsymbol{\sigma} \boldsymbol{\tilde{\sigma}})
    \end{split}    
\end{align}

\begin{figure}
    \centering
    \includegraphics[width=1.0\textwidth]{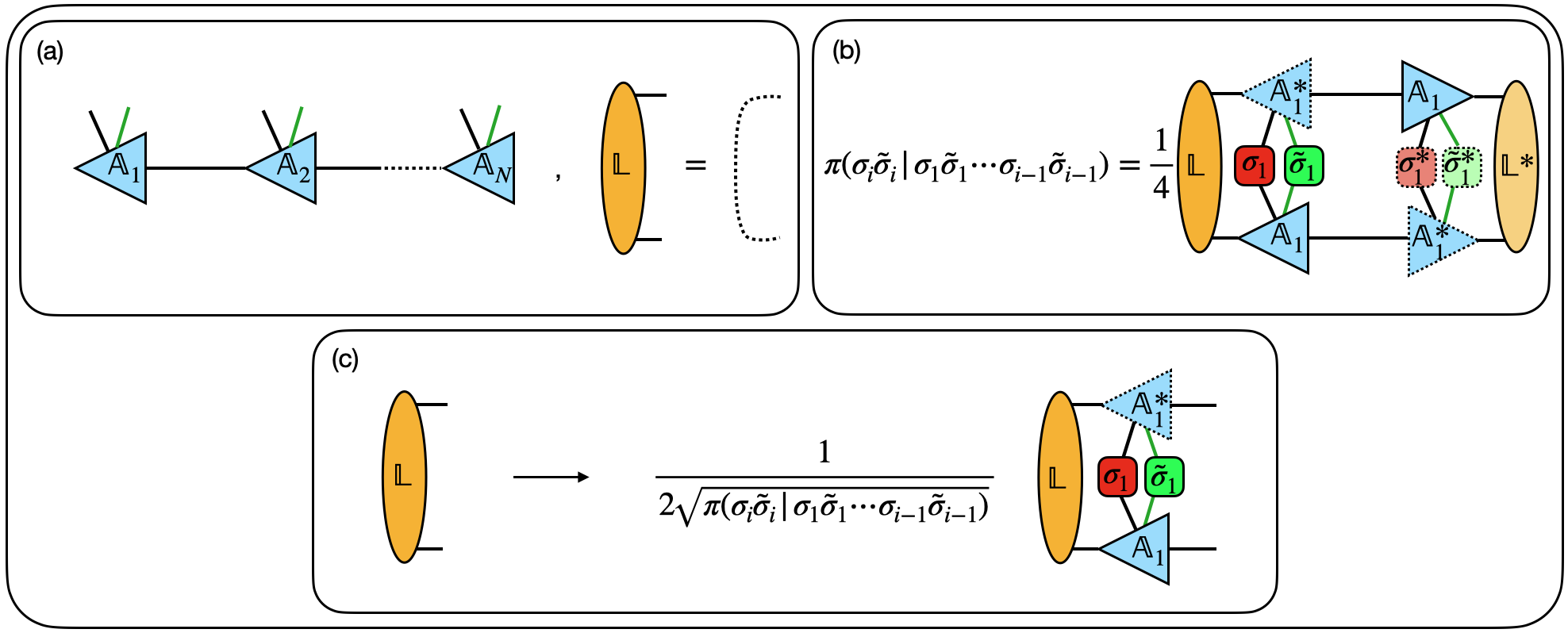}
    \caption{A diagrammatic presentation of the perfect Pauli sampling algorithm for each sampling: (a) The right normalized MPS from the TDVP and the initialization of the environment $\mathbb{L}$, (b) The calculation of the conditional probability at site $i$, (c) Update of the environment - is shown. }
    \label{fig:MPS_diag}
\end{figure}

To construct the right-hand side of Eq.~\ref{Eq:condi_prob}, we go to each $i$ to calculate the corresponding $\pi_{\rho}$. Thus, iteratively, the complete probability is constructed over a loop of length $N$. The algorithm to construct a certain Pauli string operator is discussed below:
\begin{enumerate}
    \item Initialization: we start with the right normalized MPS $\ket{X_{\beta}}$ and a local environment $\mathbb{L}$. We initialize $\mathbb{L}$ as identity $\mathbb{I}$. See Fig.~\ref{fig:MPS_diag}(a).
    \item For the first site $i=1$, we calculate the 16 possible $\pi_{\rho}(\sigma_1\tilde{\sigma}_1)$. Each of them is calculated according to the diagram shown in Fig.~\ref{fig:MPS_diag}(b) and Fig.~\ref{fig:MPS_diag_2}(a). This provides the list of the weights of all $16$ possible $\sigma_1\tilde{\sigma}_1$'s.  
    \item A certain $\sigma_1\tilde{\sigma}_1$ is randomly chosen from the $16$ with the corresponding probability $\pi_{\rho}(\sigma_1\tilde{\sigma}_1)$.
    \item Once a $\sigma_1\tilde{\sigma}_1$ is selected, $\mathbb{L}$ is updated according to the diagram shown in Fig.~\ref{fig:MPS_diag}(c). 
    \item With this updated $\mathbb{L}$ we move to the next site and repeat the steps $2-4$. This continues until $i=N$. 
\end{enumerate}

Once the loop of $i$ runs over all the sites, i.e,. up to $i=N$, one $\boldsymbol{\sigma} \boldsymbol{\tilde{\sigma}}$ is formed. To create a sample space of the length $\mathcal{N}$, we then repeat these steps $\mathcal{N}$ times. A detailed discussion of the algorithm can be found in ~\cite{LamiMario2023}.

\section{Computing SRE for mixed states}
\label{App:C}

\begin{figure}[h]
    \centering
    \includegraphics[width=1.0\textwidth]{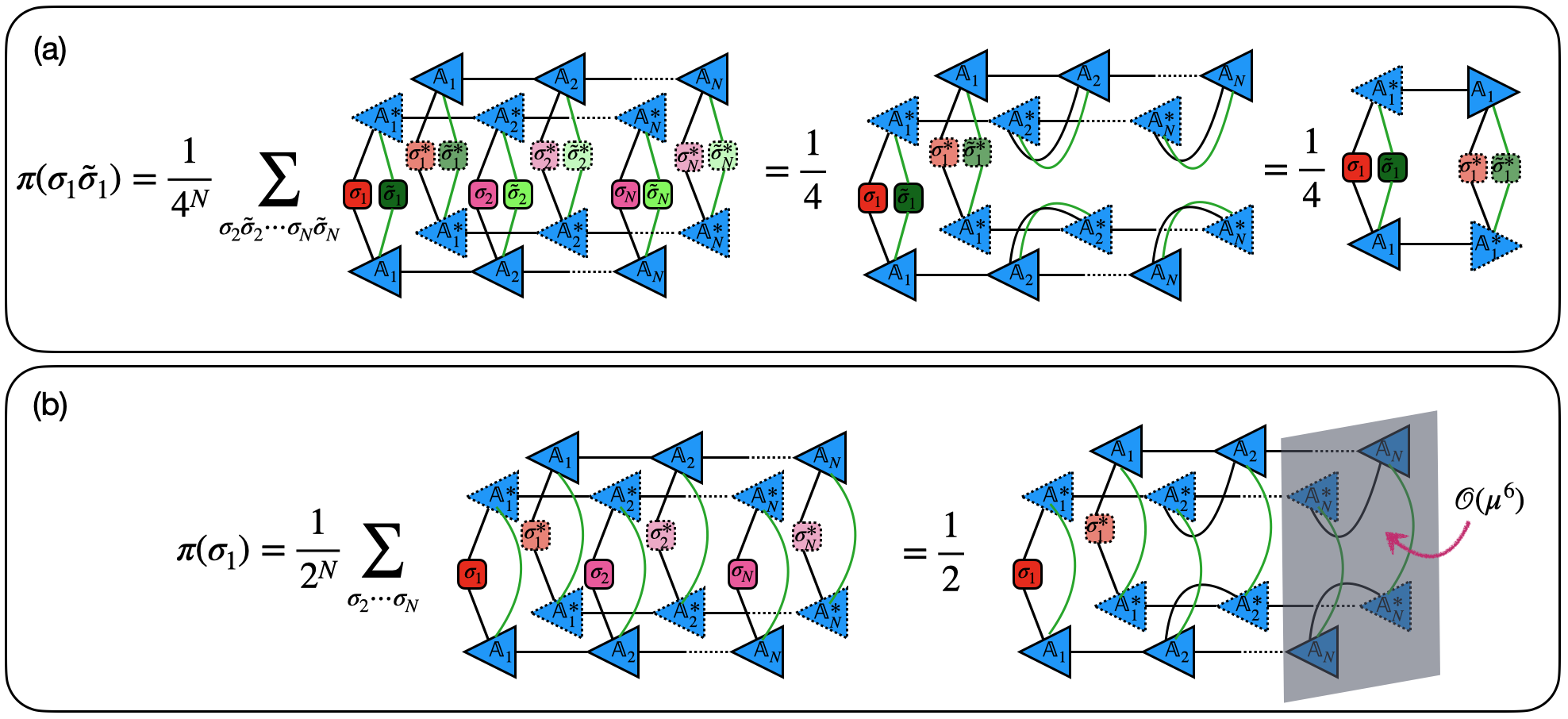}
    \caption{Two ways to deal with the auxiliary indices of the MPS: (a) Enlargement of the effective system size, which easily follows the normalization discussed in \cite{LamiMario2023}, (b) Contraction of the auxiliary indices with each other, which makes the process numerically costly. }
    \label{fig:MPS_diag_2}
\end{figure}

To calculate the SRE we use Eq.~\ref{Eq:n-SRE_MPDO} instead of Eq.~\ref{Eq:n-SRE} by enlarging the Pauli string operator and  calculating $Tr[\rho_{\beta}\boldsymbol{\sigma\tilde{\sigma}}]^{2}=\bra{X_\beta}\boldsymbol{\sigma \tilde{\sigma}}\ket{X_\beta}^2$ with  $\ket{X_\beta}=\sum_{s_1\kappa_1\cdots s_N\kappa_N }\mathbb{A}_1^{s_1\kappa_1}\cdots\mathbb{A}_1^{s_N\kappa_N}\ket{s_1\kappa_1\cdots s_N\kappa_N}$ Fig.~\ref{fig:MPS_diag}(a) while considering the Kraus indices of $\ket{X_\beta}$ as the physical indices. As mentioned in the main text, we treat $\kappa_i$ as another physical index, giving rise to an effective system size twice as large as the original one. The ideal way to deal with these extra legs is to calculate $\bra{X_\beta}\boldsymbol{\sigma}\ket{X_\beta}^2$ instead, by contracting the auxiliary leg with the corresponding conjugated one as shown in Fig.\ref{fig:MPS_diag_2}(b). But the following are the issues for this choice:
\begin{enumerate}
    \item To calculate $\bra{X_\beta}\boldsymbol{\sigma \tilde{\sigma}}\ket{X_\beta}^2$ or $\bra{X_\beta}\boldsymbol{\sigma}\ket{X_\beta}^2$  we need 4 replicas of the MPS. In the first case, the calculation of the conditional probability is straightforward, and this is clear from the depiction of the $\pi(\sigma_1\tilde{\sigma}_1)$ in Fig.~\ref{fig:MPS_diag_2}(a). Whereas in Fig.~\ref{fig:MPS_diag_2}(b) we can see the $4$ replicas get inter-connected, which makes the analysis numerically costly. For example, if we start contracting from the gray shaded part in Fig.~\ref{fig:MPS_diag_2}(b), the numerical cost is $\mu^6$ (say, $\mu$ is the bond dimension), whereas for Fig.~\ref{fig:MPS_diag_2}(a) it is $\mu^2$.
    \item Arguably, the direct contraction of the auxiliary index is not a good approximation of calculating the SRE in the finite $T$ regime. When a mixed state is represented in terms of a possible MPS, there is an extra index at each site (Fig.~\ref{fig:MPS_diag}(a)) that ensures an extra degree of freedom. For any practical purpose, this is considered just another physical leg. Therefore,  in our $T\neq0$ study, it will not be a good measure of magic if we only calculate $\bra{X_\beta}\boldsymbol{\sigma}\ket{X_\beta}^2$. 
\end{enumerate}

\section{$h_c$ from DMRG}\label{App:D}
\begin{figure}
    \centering
    \includegraphics[width=1.\textwidth]{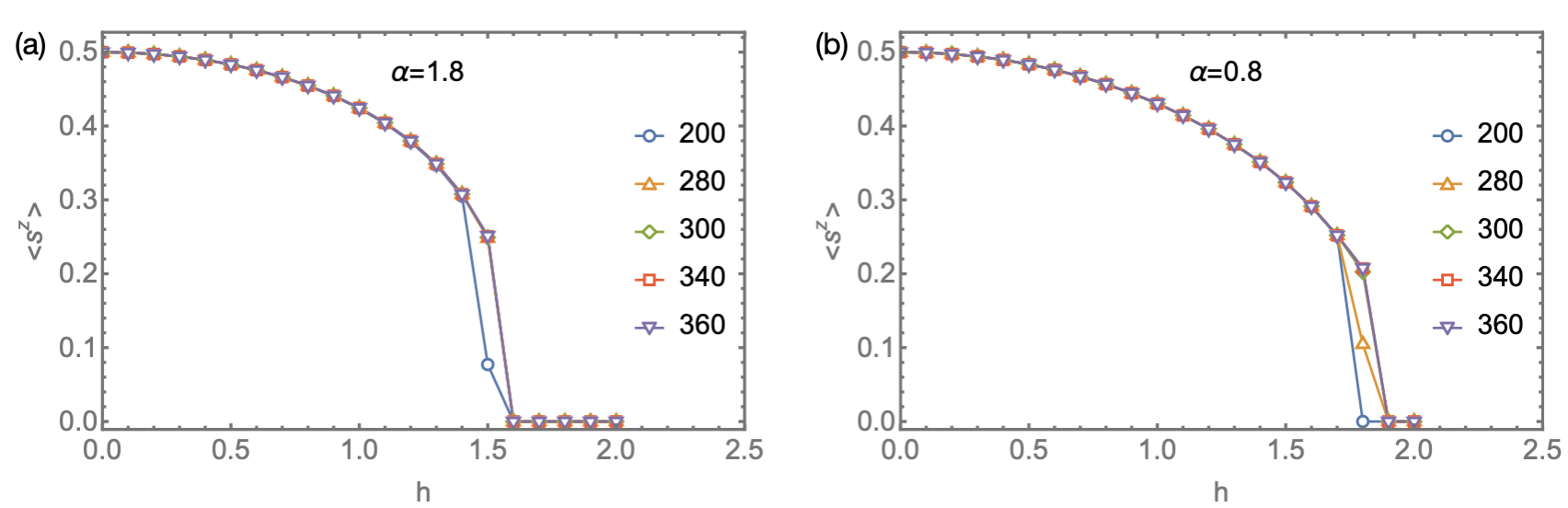}
    \caption{Evolution of the order parameter $<s^z>$ with $h$ for (a) $\alpha=1.8$, (b) $\alpha=0.8$. }
    \label{fig:DMRG}
\end{figure}

At $T=0$, to get the idea of the critical field $h_c$, we do a preliminary investigation of the quantum criticality by performing a DMRG calculation using the \textit{itensor} library of Julia \cite{itensor1,itensor2}. The non-zero to zero transition of the magnetization, $<s^z>$, indicates the quantum critical point here. In Fig.~\ref{fig:DMRG} we show how $<s^z>$ evolves as a function of $h$, for $N=200,280,300,340,360$ for each of  $\alpha=$ $1.8$ and $0.8$. We find $h_c\approx1.55$ for $\alpha=1.8$ , whereas $h_c\approx1.85$ for $\alpha=0.8$. In our discussion, the only nonzero $h$ value we consider is $h=0.3$, and therefore, for both cases, this field value is far from the quantum critical point. This is the reason why the shift in $\beta_c$ is very less as $h$ varies from $0.0$ to $0.3$ for both values of $\alpha$ ( see Table.\ref{tab:crititable} ).

\newpage

\bibliography{SciPost_Example_BiBTeX_File.bib}
\end{document}